\documentclass[12pt,useAMS,usenatbib,nomarkers,referee]{biom}

\usepackage[T1]{fontenc}
\usepackage[utf8]{inputenc}
\usepackage{lmodern}
\usepackage{textcomp}

\usepackage{fancyhdr}
\usepackage{amsmath,amssymb,bm,dsfont,mathrsfs}
\usepackage[mathscr]{euscript}
\allowdisplaybreaks   
\usepackage{thmtools,thm-restate}
\usepackage{bm}
\usepackage{graphicx}
\graphicspath{{./figs/}}
\usepackage{xcolor}
\usepackage{float}
\usepackage{subcaption}   

\usepackage{setspace}
\usepackage{enumitem}
\usepackage{titlesec}
\titlespacing*{\section}{0pt}{.25\baselineskip}{.1\baselineskip}
\titlespacing*{\subsection}{0pt}{.25\baselineskip}{.1\baselineskip}
\titlespacing*{\subsubsection}{0pt}{0.25em}{0.25em}

\usepackage{booktabs}
\usepackage{multirow}
\usepackage{tabularx}
\usepackage{threeparttable}
\usepackage{pdflscape}

\makeatletter
\@ifundefined{thead}{}{}
\makeatother
\usepackage{makecell}

\usepackage{siunitx}
\newcolumntype{L}{>{\raggedright\arraybackslash}X}
\newcolumntype{C}{>{\centering\arraybackslash}m{0.9cm}} 
\newcolumntype{A}{S[table-format=1.3]} 
\newcolumntype{O}{S[table-format=1.3]} 
\newcolumntype{R}{S[table-format=1.2]} 

\newcommand{\best}[1]{\emph{#1}}

\usepackage{authblk}
\usepackage{multicol}
\usepackage{comment}
\usepackage[normalem]{ulem} 
\usepackage{url}
\usepackage{tikz}
\usetikzlibrary{arrows.meta, positioning, shapes.geometric}

\usepackage[hidelinks]{hyperref}

\usepackage{orcidlink}
\usepackage{algorithm}
\usepackage{algpseudocode} 

\algrenewcommand\algorithmicrequire{\textbf{Input:}}
\algrenewcommand\algorithmicensure{\textbf{Output:}}

\title{Forecasting Multivariate Time Series under Predictive Heterogeneity: A Validation-Driven Clustering Framework}

  \author{Ziling Ma\,\orcidlink{0009-0006-0323-6467}\textsuperscript{1}, Ángel López-Oriona\,\orcidlink{0000-0003-1456-7342}\textsuperscript{1}, Hernando Ombao\,\orcidlink{0000-0001-7020-8091}\textsuperscript{1}, Ying Sun\,\orcidlink{0000-0001-6703-4270}\textsuperscript{1}
 \thanks{King Abdullah University of Science and Technology (KAUST), Computer, Electrical and Mathematical Sciences and Engineering (CEMSE)
 Division. Thuwal 23955-6900, Saudi Arabia. 
 Correspondence:
  ziling.ma@kaust.edu.sa} } 

\begin{document}
\pagerange{\pageref{firstpage}--\pageref{lastpage}} \pubyear{2024}

\label{firstpage}


\begin{abstract}
 
We study adaptive pooling under predictive heterogeneity in high-dimensional multivariate time series forecasting, where global models improve statistical efficiency but may fail to capture heterogeneous predictive structure, while naive specialization can induce negative transfer. We formulate adaptive pooling as a statistical decision problem and propose a validation-driven framework that determines when and how specialization should be applied. Rather than grouping series based on representation similarity, we define partitions through out-of-sample predictive performance, thereby aligning data organization with predictive risk, defined as expected out-of-sample loss and approximated via validation error. Cluster assignments are iteratively updated using validation losses for both point (Huber) and probabilistic (pinball) forecasting, improving robustness to heavy-tailed errors and local anomalies. To ensure reliability, we introduce a leakage-free fallback mechanism that reverts to a global model whenever specialization fails to improve validation performance, providing a safeguard against performance degradation under a strict training-validation-test protocol. Experiments on large-scale traffic datasets demonstrate consistent improvements over strong baselines while avoiding degradation when heterogeneity is weak. Overall, the proposed framework provides a principled and practically reliable approach to adaptive pooling in high-dimensional forecasting problems.

\end{abstract}


\maketitle
\clearpage
\pagenumbering{arabic}
\setcounter{page}{2}
\pagestyle{plain}

\section{Introduction}
\label{sec:intro}
Forecasting high-dimensional multivariate time series (MTS) is a central problem in modern statistical learning, with applications in transportation systems, energy networks, finance, and large-scale sensor monitoring. In such settings, each observation consists of many correlated components evolving over time \citep{kilian2006new}, often exhibiting heterogeneous dynamics, nonstationarity, and occasional anomalies. A fundamental challenge is to construct forecasting models that balance statistical efficiency with heterogeneity across observational units.

A recent dominant paradigm is global modeling, in which a single forecasting model is trained by pooling information across multiple series. Such approaches can substantially improve statistical efficiency and scalability in high-dimensional settings \citep{montero2021principles}. However, full pooling implicitly assumes homogeneous predictive mechanisms. When this assumption fails, global models may underfit heterogeneous dynamics, while fully local models suffer from high variance. Moreover, naive specialization through clustering may induce negative transfer, where inappropriate grouping leads to worse predictive performance than a global model. This raises a fundamental question: \emph{How should one balance global pooling and local specialization under predictive heterogeneity while avoiding negative transfer?}

To ground this question, we consider two widely studied traffic datasets, PEMS-BAY and PEMS-SF, derived from the California Performance Measurement System (PEMS), which serve as canonical benchmarks for high-dimensional MTS forecasting \citep{lee2021empirical,chattopadhyay2024context}. Both datasets exhibit substantial heterogeneity across observational units, arising from variations in traffic regimes, sensor behavior, and temporal dynamics.

Classical approaches introduce intermediate regimes via clustering, partitioning series based on similarity in feature space, latent representations, or reconstruction criteria \citep{maharaj2019time,li2019multivariate,tiano2021featts,ma2026robcpca}. While such methods provide a compromise between global and local modeling, they rely on representation-driven similarity, which need not align with the predictive objective and may induce negative transfer. 

Recent work has explored clustering based on predictive accuracy \citep{lopez2025time, lopez2026forecasting}, yielding partitions better aligned with forecasting performance. However, these approaches focus primarily on univariate time series and relatively simple model classes, and do not explicitly address robustness or specialization decisions in a leakage-free manner. In contrast, we develop a validation-driven framework for high-dimensional MTS that leverages flexible nonlinear models, incorporates robust loss functions, and introduces a leakage-free fallback mechanism to mitigate negative transfer.

In this paper, we treat clustering as a decision variable in predictive risk minimization rather than as a preprocessing step. We propose a validation-driven framework for adaptive pooling in high-dimensional MTS, in which grouping and model selection are determined by out-of-sample predictive performance. Specifically, partitions are defined through validation loss, thereby aligning specialization directly with predictive risk in the original observation space.

The framework enforces a strict separation of training, validation, and test periods, denoted as TRAIN/VAL/TEST. Model estimation is performed on TRAIN, while grouping decisions are driven exclusively by validation performance, ensuring a leakage-free protocol. Cluster assignments are iteratively updated using forecasting losses for both point and probabilistic settings, yielding an approximate minimization of predictive risk over both partitions and model parameters. To prevent over-specialization, we introduce a validation-based fallback-to-global mechanism: if a cluster-specific model fails to outperform the global model on validation data, its members are routed to the global forecaster. This induces a safeguard against performance degradation, ensuring that specialization is applied only when it improves predictive performance relative to the pooled model.

Beyond improving predictive accuracy, the proposed framework provides insight into predictive heterogeneity by identifying which series benefit from specialization and which are better served by global modeling.

A key practical implication is its applicability to new time series. Given a newly observed MTS from the same context, the framework uses an initial observed segment to assign the series to the model that minimizes short-horizon predictive loss. The selected model is then used to forecast future values of the same series, yielding adaptive and robust forecasting at deployment.

A key practical implication is its applicability to new time series. Given a newly observed MTS, the framework uses an initial observed segment to assign the series to the model that minimizes short-horizon predictive loss, with fallback to the GLOBAL model when specialization is not beneficial. The selected model is then used to forecast future values of the same series, yielding adaptive and robust forecasting at deployment.

The main contributions of this paper are as follows. 
First, we formulate clustering for MTS forecasting as a decision problem in predictive risk, in which partitions are defined through out-of-sample performance rather than representation similarity. 
Second, we develop a validation-driven, model-agnostic framework that explicitly separates estimation (on TRAIN) from decision-making (on VAL), thereby ensuring a leakage-free protocol for clustering, model selection, and evaluation. 
Third, we introduce a validation-based routing rule with fallback to a pooled model, which controls negative transfer by applying specialization only when it improves predictive performance. 
Fourth, we incorporate robust loss functions, including the Huber and pinball losses, to enhance stability under heavy-tailed errors and local anomalies. 
Finally, empirical results on high-dimensional traffic datasets demonstrate consistent improvements over strong baselines while maintaining robustness when predictive heterogeneity is weak.

The remainder of the paper is organized as follows. Section~\ref{sec:framework} presents the proposed framework. Section~\ref{sec:point} develops the point forecasting formulation, while Section~\ref{sec:prob_forecast} extends it to probabilistic forecasting. Section~\ref{sec:real_data} reports the empirical results, and Section~\ref{sec:conclusion} concludes.

\section{A clustering framework for MTS based on forecasting accuracy}
\label{sec:framework}

This section outlines the proposed framework, including TRAIN-based prototype fitting, VAL-based reassignment and fallback decisions, and single-use TEST evaluation.

\subsection{Framework overview}
\label{sec:overview}

We propose a leakage-free clustering framework that partitions MTS into non-overlapping clusters according to their forecastability. Forecastability is defined through out-of-sample predictive performance under a specified forecasting model class, rather than through similarity-based criteria. The central principle is to separate (i) prototype training on TRAIN from (ii) validation-based reassignment and model selection on VAL, while reserving TEST exclusively for final evaluation.

The procedure begins by initializing the $N$ MTS into $K$ clusters, for example via random balanced assignment or feature-based initialization. A single GLOBAL forecasting model is then trained on TRAIN using all $N$ series. This model serves both as a stable pooled baseline and as a reference for warm-starting the cluster-specific prototypes. Given the current cluster assignments, $K$ cluster-specific forecasting prototypes are subsequently fitted on TRAIN.

Cluster labels are updated by assigning each series to the prototype that minimizes its validation forecasting loss, yielding a forecastability-driven reassignment that is free of test leakage. Prototype fitting (TRAIN) and validation-based reassignment (VAL) are alternated until the assignments stabilize or a maximum number of outer iterations is reached.

To guard against over-specialization, we introduce a validation-based fallback-to-GLOBAL safeguard. A cluster is declared non-specializable if its average validation loss under its specialized prototype exceeds that of the GLOBAL model evaluated on the same members. In this case, its members are routed to the GLOBAL forecaster at evaluation time. Importantly, the fallback decision is determined on VAL and \emph{frozen} prior to the final refit. TEST is used only once.

\begin{figure}[htbp]
\centering
\resizebox{\linewidth}{!}{%
\begin{tikzpicture}[
    node distance=1.15cm and 1.15cm,
    every node/.style={font=\small, align=center},
    startstop/.style={
        rectangle, rounded corners,
        draw,
        text width=3.0cm,
        minimum height=1.05cm
    },
    process/.style={
        rectangle,
        draw,
        text width=3.35cm,
        minimum height=1.10cm
    },
    decision/.style={
        diamond,
        draw,
        text width=2.2cm,
        aspect=2.2,
        inner sep=1pt
    },
    arrow/.style={->, thick}
]

\node (split)   [startstop] {Leakage-free split\\TRAIN / VAL / TEST};
\node (init)    [process, right=of split] {Initialize $N$ series\\into $K$ clusters};
\node (global)  [process, right=of init] {Train GLOBAL forecaster\\on TRAIN\\(baseline + anchor)};
\node (proto)   [process, right=of global] {Train $K$ cluster prototypes\\on TRAIN\\(warm start + $\ell_2$-SP)};
\node (assign)  [process, right=of proto] {VAL-driven reassignment (VAL loss):\\assign each series to\\best-forecasting prototype};
\node (conv)    [decision, right=1.6cm of assign] {Assignments\\converged?};

\node (freeze)  [process, below=2.1cm of conv] {Compute fallback flags on VAL\\\& {freeze} decisions:\\cluster worse than GLOBAL $\Rightarrow$\\non-specializable (route to GLOBAL)};
\node (refit)   [process, left=of freeze] {Refit GLOBAL + all prototypes\\on TRAIN+VAL\\(flags fixed from VAL)};
\node (test)    [startstop, left=of refit] {Single-use TEST evaluation:\\use GLOBAL if flagged,\\else use prototype};
\node (loss)    [process, left=of test] {Forecasting loss:\\Huber (point)\\or Pinball (prob.)};

\draw [arrow] (split) -- (init);
\draw [arrow] (init) -- (global);
\draw [arrow] (global) -- (proto);
\draw [arrow] (proto) -- (assign);
\draw [arrow] (assign) -- (conv);

\draw [arrow] (conv.north) -- ++(0,0.85)
node[above]{no (iterate)}
-| (proto.north);

\draw [arrow] (conv) -- node[right]{yes} (freeze);

\draw [arrow] (freeze) -- (refit);
\draw [arrow] (refit) -- (test);
\draw [arrow] (test) -- (loss);

\end{tikzpicture}%
}
\caption{Schematic of the proposed validation-driven clustering framework for adaptive pooling under predictive heterogeneity.
}

\label{flowchart}
\end{figure}
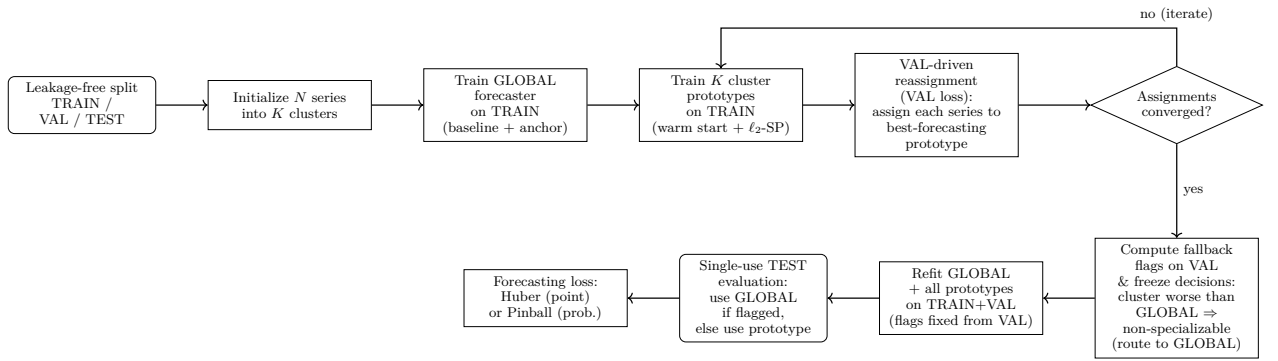

The same leakage-free protocol applies to both point and probabilistic forecasting, differing only in the choice of forecasting loss and prediction target (point versus quantiles). Figure~\ref{flowchart} provides a concise overview of the proposed framework.

\subsection{Data preparation}
\label{sec:data_prep}

We describe the leakage-free data preparation procedure, including the strict time split and TRAIN-only preprocessing.

\subsubsection{Data, notation, and strict time split}
\label{sec:split}

Let $\{\mathbf{X}_i\}_{i=1}^N$ denote $N$ MTS, where $\mathbf{X}_i=(\mathbf{x}_{i,1},\ldots,\mathbf{x}_{i,T})$ with $\mathbf{x}_{i,t}\in\mathbb{R}^P$ and component $x_{i,t,p}$ for $p=1,\ldots,P$. Fix integers $T_{\mathrm{tr}},T_{\mathrm{va}},T_{\mathrm{te}}>0$ such that $T_{\mathrm{tr}}+T_{\mathrm{va}}+T_{\mathrm{te}}=T$. We adopt a strict chronological split:
\[
\mathbf{X}_i^{\mathrm{tr}}=(\mathbf{x}_{i,1:T_{\mathrm{tr}}}),\quad
\mathbf{X}_i^{\mathrm{va}}=(\mathbf{x}_{i,T_{\mathrm{tr}}+1:T_{\mathrm{tr}}+T_{\mathrm{va}}}),\quad
\mathbf{X}_i^{\mathrm{te}}=(\mathbf{x}_{i,T_{\mathrm{tr}}+T_{\mathrm{va}}+1:T}).
\]
TRAIN is used for model fitting, VAL for reassignment and model selection, and TEST is used once for final evaluation. All preprocessing statistics are computed on TRAIN only and then applied to VAL and TEST, ensuring a leakage-free protocol.

\subsubsection{Imputation and standardization}
\label{sec:impute_scale}

Missing values are imputed using TRAIN-only statistics. For each component $p$, define $\mathcal{I}_p^{\mathrm{tr}}=\{(i,t): 1\le i\le N,\ 1\le t\le T_{\mathrm{tr}},\ x_{i,t,p}\ \text{observed}\}$ and compute $\mu_p = |\mathcal{I}_p^{\mathrm{tr}}|^{-1}\sum_{(i,t)\in\mathcal{I}_p^{\mathrm{tr}}} x_{i,t,p}$. Missing entries in all splits are replaced by $\mu_p$ (a median-based variant can be used analogously).

We then standardize each component using TRAIN statistics: $\sigma_p = \big(|\mathcal{I}_p^{\mathrm{tr}}|^{-1}\sum_{(i,t)\in\mathcal{I}_p^{\mathrm{tr}}}(x_{i,t,p}-\mu_p)^2 + \varepsilon\big)^{1/2}$ with $\varepsilon>0$ for numerical stability, and apply the affine transform $\widetilde{x}_{i,t,p} = (x_{i,t,p}-\mu_p)/\sigma_p$ for $(i,t)\in \mathrm{tr},\mathrm{va},\mathrm{te}$. This ensures that no information from VAL or TEST is used in preprocessing, preserving leakage-free evaluation.

\subsection{Forecasting setup and robust losses}
\label{sec:setup_losses}

This section introduces the forecasting setup and the loss functions used to define forecastability within the proposed framework.

\subsubsection{Horizon-wise losses, forecasting task, and objective formulation}
\label{sec:horizons}

Let $\mathcal{H}\subset\mathbb{N}$ denote the set of forecast horizons considered in this paper. For each $h\in\mathcal{H}$, we evaluate the loss for predicting $\mathbf{x}_{i,t+h}$ using information available up to time $t$, that is, for an $h$-step-ahead forecasting task.

Fix a window length $w$. We predict $\mathbf{x}_{i,t+h}$ from the most recent window $\mathbf{x}_{i,t-w+1:t}$. Let $E_\phi$ denote an encoder mapping $\mathbb{R}^P$ to a lower-dimensional representation, let $f_\theta$ denote a sequence forecaster operating in the encoded space, and let $D_\psi$ denote a decoder reconstructing predictions back to $\mathbb{R}^P$. The resulting $h$-step-ahead prediction in the original space is
\begin{equation}
\label{eq:pred_point_generic}
\widehat{\mathbf{x}}_{i,t+h}(\theta,\phi,\psi)
=
D_\psi\!\left(f_\theta\!\big(E_\phi(\mathbf{x}_{i,t-w+1:t})\big)\right)\in\mathbb{R}^P.
\end{equation}

Let $s\in\{\mathrm{tr},\mathrm{va},\mathrm{te}\}$ index the data split, and let $\mathcal{W}_i^s(h)$ denote the set of valid window end-times for series $i$ at horizon $h$. Given a pointwise loss $\ell(\cdot,\cdot)$, define the split-specific mean forecasting loss
\begin{equation}
\label{eq:split_loss_generic}
\mathcal{L}_i^s(h;\theta,\phi,\psi)
=
\frac{1}{|\mathcal{W}_i^s(h)|}
\sum_{t\in\mathcal{W}_i^s(h)}
\ell\!\left(\widehat{\mathbf{x}}_{i,t+h}(\theta,\phi,\psi),\,\mathbf{x}_{i,t+h}\right).
\end{equation}

The proposed clustering framework can be interpreted as an approximate minimization of a validation-based predictive risk over both cluster assignments and forecasting parameters.

Let $\mathbf{c}=(c_1,\ldots,c_N)$ denote the cluster assignment vector, where $c_i\in\{1,\ldots,K\}$. Let $\Theta_g$ denote the parameters of the GLOBAL model, and let $\Theta_k$ denote the parameters of prototype $k$, $k=1,\ldots,K$. Define $I_k=\{i:c_i=k\}$.

To formalize the validation-based fallback mechanism, define the set of \emph{specializable clusters} as
\begin{equation}
\label{eq:specializable_set}
\mathcal{S}(\mathbf{c},\Theta_g,\Theta_1,\ldots,\Theta_K)
=
\left\{
k:
\frac{1}{|I_k|}
\sum_{i\in I_k}
\mathcal{L}_i^{\mathrm{va}}(1;\Theta_k)
\le
\frac{1}{|I_k|}
\sum_{i\in I_k}
\mathcal{L}_i^{\mathrm{va}}(1;\Theta_g)
\right\}.
\end{equation}

For a given assignment $\mathbf{c}$, the routed validation loss for series $i$ is then defined as
\begin{equation}
\label{eq:routed_loss}
\widetilde{\mathcal{L}}_i^{\mathrm{va}}(1;\mathbf{c},\Theta_g,\Theta_1,\ldots,\Theta_K)
=
\begin{cases}
\mathcal{L}_i^{\mathrm{va}}(1;\Theta_{c_i}), & \text{if } c_i \in \mathcal{S}(\mathbf{c},\Theta_g,\Theta_1,\ldots,\Theta_K), \\
\mathcal{L}_i^{\mathrm{va}}(1;\Theta_g), & \text{otherwise}.
\end{cases}
\end{equation}

We then define the overall validation risk as
\begin{equation}
\label{eq:routed_val_risk}
\mathcal{R}_{\mathrm{va}}(\mathbf{c},\Theta_g,\Theta_1,\ldots,\Theta_K)
=
\frac{1}{N}\sum_{i=1}^N
\widetilde{\mathcal{L}}_i^{\mathrm{va}}(1;\mathbf{c},\Theta_g,\Theta_1,\ldots,\Theta_K).
\end{equation}

The clustering problem can be formulated as
\begin{equation}
\label{eq:joint_objective}
\left\{
\begin{aligned}
\min_{\mathbf{c},\,\Theta_g,\,\Theta_1,\ldots,\Theta_K}
\quad & \mathcal{R}_{\mathrm{va}}(\mathbf{c},\Theta_g,\Theta_1,\ldots,\Theta_K)  \\
\text{s.t.}\quad
& \Theta_g \in \arg\min_{\Theta} \sum_{i=1}^N \mathcal{L}_i^{\mathrm{tr}}(1;\Theta), \\
& \Theta_k \in \arg\min_{\Theta} \sum_{i:c_i=k} \mathcal{L}_i^{\mathrm{tr}}(1;\Theta),
\quad k=1,\ldots,K.
\end{aligned}
\right.
\end{equation}

The constraints in Equation \eqref{eq:joint_objective} define the model parameters $\Theta_g$ and $\{\Theta_k\}_{k=1}^K$ as solutions to training-risk minimization problems conditional on the assignments, so that the parameters are implicitly determined by $\mathbf{c}$. As a result, the joint problem can be viewed as an optimization over the partition, with parameters acting as induced quantities. This separation also enforces a leakage-free design, where TRAIN data is used for parameter estimation and validation data governs clustering decisions.

This formulation makes explicit that clustering is treated as a decision variable in predictive risk minimization, rather than as a preprocessing step based on representation similarity. In practice, Equation \eqref{eq:joint_objective} is not solved jointly. instead, we adopt an alternating procedure in which (i) given assignments, prototypes are estimated on TRAIN, and (ii) given parameters, assignments are updated by minimizing validation losses. The fallback safeguard is then applied to prevent negative transfer, and its decisions are fixed prior to final refitting.

In practical implementations, prototype estimation may incorporate regularization or warm-start strategies; these are introduced in Section~\ref{sec:point} for the specific forecasting instantiation. For stability and computational efficiency, model parameters are estimated using the one-step objective ($h=1$), while multi-step forecastability is assessed on validation data via recursive rollout over user-specified horizon sets.

\subsubsection{Point forecasting loss: Huber}
\label{sec:huber}

For point forecasting, we use the component-wise Huber (smooth-$\ell_1$) loss. Given the $h$-step-ahead prediction $\widehat{\mathbf{x}}_{i,t+h}\in\mathbb{R}^P$ and the observation $\mathbf{x}_{i,t+h}\in\mathbb{R}^P$, define
\begin{equation}
\label{eq:huber_loss}
\ell_\delta(\widehat{\mathbf{x}}_{i,t+h},\mathbf{x}_{i,t+h})
=
\frac{1}{P}\sum_{p=1}^P
\begin{cases}
\frac{1}{2}e_{i,t+h,p}^2, & |e_{i,t+h,p}|\le \delta, \\[3pt]
\delta |e_{i,t+h,p}|-\frac{1}{2}\delta^2, & |e_{i,t+h,p}|>\delta,
\end{cases}
\end{equation}
where $e_{i,t+h,p}=\widehat{x}_{i,t+h,p}-x_{i,t+h,p}$. The Huber loss is quadratic near zero and linear in the tails, combining sensitivity to small residuals with reduced influence of large forecast errors \citep{huber1992robust}. It is therefore more robust than squared loss under heavy-tailed errors or local anomalies, while remaining smoother and easier to optimize than absolute loss \citep{dwaram2022crop,xing2022forecasting}. The parameter $\delta>0$ controls the transition between the quadratic and linear regimes\footnote{We set $\delta=1$, a standard and slightly conservative choice for the Huber loss after standardization, relative to classical values around $1$--$1.345$ \citep{lambert2011robust}.}.

\subsubsection{Probabilistic forecasting loss: pinball}
\label{sec:pinball}

For probabilistic forecasting, we predict conditional quantiles of each component of $\mathbf{x}_{i,t+h}$. Let $\widehat{x}_{i,t+h,p}^{(q)}$ denote the predicted $q$-quantile for component $p$, where $q\in(0,1)$. The pinball loss for a single component and quantile level is
\begin{equation}
\label{eq:pinball_scalar}
\rho_q\!\left(x_{i,t+h,p}-\widehat{x}_{i,t+h,p}^{(q)}\right)
=
\begin{cases}
q\left(x_{i,t+h,p}-\widehat{x}_{i,t+h,p}^{(q)}\right), & x_{i,t+h,p}\ge \widehat{x}_{i,t+h,p}^{(q)}, \\[3pt]
(q-1)\left(x_{i,t+h,p}-\widehat{x}_{i,t+h,p}^{(q)}\right), & x_{i,t+h,p}< \widehat{x}_{i,t+h,p}^{(q)}.
\end{cases}
\end{equation}
Equivalently, if $u=x_{i,t+h,p}-\widehat{x}_{i,t+h,p}^{(q)}$, then $\rho_q(u)=u\bigl(q-\mathbf{1}\{u<0\}\bigr)$.

The pinball loss is the canonical loss for quantile regression: its population minimizer is the $q$-quantile of the target distribution. If $Y$ is a real-valued random variable with distribution function $F_Y$, then
\[
Q_Y(q)=\inf\{y\in\mathbb{R}:F_Y(y)\ge q\}
\]
minimizes $\mathbb{E}\{\rho_q(Y-a)\}$ over $a\in\mathbb{R}$ \citep{koenker1978regression}. Accordingly, minimizing pinball loss yields forecasts targeted at conditional quantiles rather than conditional means. Its asymmetry allows under- and over-prediction to be weighted differently according to $q$, making it well suited for probabilistic forecasting and uncertainty quantification, particularly when multiple quantiles are used to construct prediction intervals \citep{wang2019probabilistic,cui2025probabilistic}. For multivariate observations and a quantile grid $\mathcal{Q}$, we average the loss over components and quantile levels.

\subsection{Complete leakage-free procedure}
\label{sec:algorithm}

The Supplementary Material (Algorithm \ref{alg:forecast_clustering}) provides the full algorithm for the leakage-free protocol, including TRAIN-based fitting, VAL-based reassignment and fallback decisions, final refitting, and single-use TEST evaluation. The framework is model-agnostic. The forecasting architecture used in our experiments is described separately in Section~\ref{sec:arch}.

From a statistical perspective, the framework can be viewed as an approximate minimization of validation-based predictive risk over both model parameters and partition structure. By separating estimation on TRAIN from decision-making on VAL, it avoids optimistic bias and provides a principled criterion for specialization. The fallback mechanism ensures that specialization is applied only when it reduces validation risk relative to the pooled GLOBAL model. Consequently, the method implements a validation-based no-regret decision rule at the level of cluster assignments, preventing degradation in out-of-sample performance under the adopted protocol.

The proposed framework also extends naturally to forecasting newly observed series
arising under the same structural regime. Given an initial observed segment of a new MTS, short-horizon predictive loss can be evaluated under the GLOBAL model and each cluster-specific prototype using the same loss function as in validation. The new series is then assigned to the model that attains the smallest loss. Consistent with the validation-driven procedure, fallback to the GLOBAL model is applied whenever no specialized prototype improves upon the pooled forecaster. The selected model is subsequently used to forecast future observations of the series, providing an adaptive deployment rule without retraining.

\subsection{Forecaster architecture used in experiments}
\label{sec:arch}

Although the proposed framework is model-agnostic, all experiments implement the forecasting map in Equation \eqref{eq:pred_point_generic} using a lightweight neural architecture consisting of three components: (i) a learned linear mixture layer, (ii) a one-layer gated recurrent unit (GRU), and (iii) a linear prediction head.

Given input \(\mathbf{x}_{i,t}\in\mathbb{R}^P\), the mixture layer maps it to a lower-dimensional representation
\[
\mathbf{z}_{i,t} = B \mathbf{x}_{i,t}, \qquad B\in\mathbb{R}^{r\times P}, \quad r \ll P,
\]
where \(r\) is the latent dimension. This layer acts as a learned linear dimension-reduction step, reducing computational cost and mitigating overparameterization in high-dimensional settings.

The latent sequence \(\{\mathbf{z}_{i,t}\}\) is then processed by a one-layer GRU, which models temporal dependence through gated updates and alleviates vanishing-gradient issues. GRUs have been shown to provide competitive performance relative to more complex architectures while remaining computationally efficient \citep{chung2014empirical}. Similar recurrent architectures have also been used successfully in probabilistic forecasting \citep{salinas2020deepar}.

The final hidden state is mapped back to the latent space by a linear prediction head, and the forecast is reconstructed in the original \(P\)-dimensional space using the transpose of \(B\). The resulting architecture combines dimensionality reduction with sequence modeling, enabling efficient learning in high-dimensional settings.

This design is intentionally lightweight. Because the framework repeatedly refits GLOBAL and cluster-specific models across iterations, seeds, and candidate values of \(K\), a simple architecture helps maintain computational efficiency and ensures that performance gains can be attributed to the proposed forecastability-driven clustering procedure rather than architectural complexity.

\section{Point forecasting instantiation}
\label{sec:point}

We instantiate the general framework of Section~\ref{sec:framework} for point forecasting. The clustering, reassignment, fallback, and model-selection protocol remains unchanged; only the forecasting target and loss function are specialized. We use the Huber loss $\ell_\delta$ in Equation \eqref{eq:huber_loss} for validation-driven reassignment and model selection.

\subsection{GLOBAL forecaster and warm-started prototypes}
\label{sec:point_models}

Although losses are defined for all horizons \(h\in\mathcal{H}\), we distinguish between the horizon used for parameter estimation and those used for validation-based reassignment and cluster-number selection.\footnote{This setup assumes a form of forecastability stability across training and validation: although the losses may differ, the relative ordering of candidate assignments is approximately preserved.} We train the GLOBAL model and all prototypes using the one-step objective (\(h=1\)), which is computationally stable and data-efficient within the iterative refit--reassignment loop.

Multi-step forecastability is incorporated on validation data via the assignment horizon set \(\mathcal{H}_a\subseteq\mathcal{H}\), used to average validation losses over \(h\in\mathcal{H}_a\). In our implementation, the number of clusters \(K\) is selected using routed validation Huber loss at \(h=1\). For \(h>1\), validation and test losses are evaluated via recursive rollout.

Fit a GLOBAL model on TRAIN:
\begin{equation}
\label{eq:global_fit}
(\theta_g,\phi,\psi)
=
\arg\min_{\theta,\phi,\psi}\sum_{i=1}^N \mathcal{L}_{i}^{\mathrm{tr}}(1;\theta,\phi,\psi),
\end{equation}
with $\ell=\ell_\delta$. The resulting model serves as both a pooled baseline and a warm-start anchor.

Let $c_i\in\{1,\ldots,K\}$ denote cluster assignments. We share $(\phi,\psi)$ across clusters and specialize only $\theta$. Given assignments, prototype $k$ is estimated via
\begin{equation}
\label{eq:l2sp_obj}
\theta_k
=
\arg\min_{\theta}
\sum_{i:c_i=k}\mathcal{L}_{i}^{\mathrm{tr}}(1;\theta,\phi,\psi)
+
\eta\,\|\theta-\theta_g\|_2^2,
\end{equation}
where $\eta\ge 0$ controls warm-start ($\ell_2$-SP) regularization \citep{xuhong2018explicit}.

\subsection{VAL-driven reassignment and fallback}
\label{sec:point_assign}

Given prototypes $\{\theta_k\}_{k=1}^K$, assignments are updated using validation losses. The VAL cost matrix $\mathbf{C}\in\mathbb{R}^{N\times K}$ is
\begin{equation}
\label{eq:val_cost}
C_{ik}
=
\frac{1}{|\mathcal{H}_a|}
\sum_{h\in\mathcal{H}_a}
\mathcal{L}_{i}^{\mathrm{va}}(h;\theta_k,\phi,\psi),
\end{equation}
and assignments are updated as
\begin{equation}
\label{eq:assign_update}
c_i \leftarrow \arg\min_{k} C_{ik}.
\end{equation}

Prototype fitting on TRAIN and reassignment on VAL are alternated until convergence or a maximum number of iterations is reached.

After convergence, we identify clusters that fail to improve over the GLOBAL model using VAL and freeze this decision. For each cluster $k$, compare
\[
L_k^{\mathrm{clus}}
=
\frac{1}{|\{i:c_i=k\}|}\sum_{i:c_i=k}\mathcal{L}_i^{\mathrm{va}}(1;\theta_k,\phi,\psi),
\quad
L_k^{\mathrm{glob}}
=
\frac{1}{|\{i:c_i=k\}|}\sum_{i:c_i=k}\mathcal{L}_i^{\mathrm{va}}(1;\theta_g,\phi,\psi).
\]
If $L_k^{\mathrm{clus}}>L_k^{\mathrm{glob}}$, cluster $k$ is declared non-specializable and its members are routed to the GLOBAL forecaster at TEST time.

\subsection{Selecting $K$ by validation forecastability}
\label{sec:k-selection}

We evaluate candidate values $K\in\mathcal{K}$, where larger $K$ allows greater specialization at the cost of increased complexity. For each $K$ and initialization $s$, we run the full TRAIN--VAL procedure and compute the routed validation loss after reassignment and fallback.

Let $\tilde c_i$ denote the final routed assignment. We define
\begin{equation}
\label{eq:selabs}
\mathrm{SelAbs}_{\mathrm{VAL}}(K,s)
=
\frac{1}{N}
\sum_{i=1}^N
\widetilde{\mathcal{L}}_i^{\mathrm{va}}\bigl(1;\tilde{\mathbf{c}},\widehat{\Theta}_g,\widehat{\Theta}_1,\ldots,\widehat{\Theta}_K\bigr),
\end{equation}
where $\widetilde{\mathcal{L}}_i^{\mathrm{va}}$ is the routed validation loss defined in Section~\ref{sec:horizons}.

We select $(K,s)$ via the penalized criterion
\begin{equation}
\label{eq:selpen}
\mathrm{SelPen}_{\mathrm{VAL}}(K,s)
=
\mathrm{SelAbs}_{\mathrm{VAL}}(K,s)
+
\gamma\frac{K}{N},
\qquad \gamma \ge 0.
\end{equation}

In our implementation, we set $\gamma=0.05$, which provides a mild penalty and yields stable selections across $K$. Results are not sensitive to moderate variations of $\gamma$, and similar selections of $K$ and forecasting performance are obtained over a reasonable range of values.

\section{Probabilistic forecasting extension via quantile regression}
\label{sec:prob_forecast}

We extend the framework to probabilistic forecasting by predicting a grid of conditional quantiles. Let $\mathcal{Q}=\{q_1,\ldots,q_Q\}\subset(0,1)$ denote the quantile levels. For each horizon $h$, the forecaster outputs $\widehat{\mathbf{x}}_{i,t+h}^{(q)}\in\mathbb{R}^P$, and training, reassignment, and model selection are driven by the pinball loss.

\subsection{Quantile prediction and pinball training}
\label{sec:quantile_train}

For each $q\in\mathcal{Q}$, quantiles are estimated via quantile regression by minimizing the empirical pinball loss. The model produces latent representations $\widehat{\mathbf{z}}_{i,t+h}^{(q)}(\Theta)\in\mathbb{R}^{d}$, decoded as
\[
\widehat{\mathbf{x}}_{i,t+h}^{(q)}(\Theta,\phi,\psi)
=
D_\psi\!\left(\widehat{\mathbf{z}}_{i,t+h}^{(q)}(\Theta)\right).
\]
All quantiles share the encoder and recurrent backbone and are trained jointly. To ensure non-crossing quantiles, we adopt a monotone parameterization (Section~\ref{sec:noncrossing}).

Define the split-specific pinball loss as
\begin{equation}
\label{eq:pinball_multi}
\mathcal{L}^{s,\mathrm{PB}}_{i}(h;\Theta,\phi,\psi)
=
\frac{1}{|\mathcal{W}_i^{s}(h)|}
\sum_{t\in\mathcal{W}_i^{s}(h)}
\left[
\frac{1}{P|\mathcal{Q}|}\sum_{p=1}^P\sum_{q\in\mathcal{Q}}
\rho_q\!\left(
x_{i,t+h,p}-\widehat{x}_{i,t+h,p}^{(q)}(\Theta,\phi,\psi)
\right)
\right].
\end{equation}
Reassignment follows the same VAL-driven criterion with $\mathcal{L}_i^{s}$ replaced by $\mathcal{L}_i^{s,\mathrm{PB}}$, yielding
\[
C_{ik}
=
\frac{1}{|\mathcal{H}_a|}
\sum_{h\in\mathcal{H}_a}
\mathcal{L}^{\mathrm{va},\mathrm{PB}}_{i}\!\left(h;\Theta_k,\phi,\psi\right).
\]
All remaining steps are unchanged.

\subsection{Non-crossing quantiles}
\label{sec:noncrossing}

To enforce monotonicity across quantiles, we adopt a cumulative-sum parameterization. Let $\mathbf{b}_{i,t+h}\in\mathbb{R}^P$ be a base vector and $\{\boldsymbol{\Delta}^{(m)}_{i,t+h}\}_{m=2}^Q$ unconstrained increments. Define
\[
\widehat{\mathbf{z}}^{(q_1)}_{i,t+h} = \mathbf{b}_{i,t+h}, \qquad
\widehat{\mathbf{z}}^{(q_j)}_{i,t+h}
=
\mathbf{b}_{i,t+h}
+
\sum_{m=2}^{j}
\mathrm{softplus}\!\left(\boldsymbol{\Delta}^{(m)}_{i,t+h}\right), \quad j=2,\ldots,Q.
\]
Since $\mathrm{softplus}(\cdot)\ge 0$, the resulting quantiles are non-crossing by construction. Predictions in the observation space are obtained via the decoder.

\subsection{Multi-step rollout and interval diagnostics}
\label{sec:intervals}

For $h>1$, we use recursive rollout. Reassignment and model selection propagate only the median path ($q=0.5$), while interval diagnostics use $(q_\ell,0.5,q_u)$ to capture uncertainty growth.

Given intervals $[\widehat{\mathbf{x}}^{(q_\ell)}_{i,t+h},\widehat{\mathbf{x}}^{(q_u)}_{i,t+h}]$, we report coverage and mean width:
\[
\mathrm{Cov}(h)
=
\frac{1}{NP}\sum_{i=1}^N\sum_{p=1}^P
\mathbb{I}\!\left\{
x_{i,t+h,p}\in
\left[
\widehat{x}^{(q_\ell)}_{i,t+h,p},
\widehat{x}^{(q_u)}_{i,t+h,p}
\right]
\right\},
\quad
\mathrm{Wid}(h)
=
\frac{1}{NP}\sum_{i=1}^N\sum_{p=1}^P
\left(
\widehat{x}^{(q_u)}_{i,t+h,p}
-
\widehat{x}^{(q_\ell)}_{i,t+h,p}
\right).
\]

\subsection{VAL calibration and fallback}
\label{sec:calib}

We optionally calibrate intervals on VAL using a scalar factor $s(h)\ge 0$ per horizon. Let $\bm{\widehat{m}}_{i,t+h}$ denote the median prediction and $\bm{\widehat{q}}^{(\ell)}_{i,t+h}$, $\bm{\widehat{q}}^{(u)}_{i,t+h}$ the lower and upper quantiles. The calibrated bounds are
We optionally calibrate predictive intervals on VAL using a scalar inflation factor
$s(h)\ge 0$ for each forecast horizon $h$.
Let $\widehat{\bm{m}}_{i,t+h}$ denote the median prediction and
$\widehat{\bm{q}}_{i,t+h}^{(\ell)}$, $\widehat{\bm{q}}_{i,t+h}^{(u)}$ the lower and upper
quantiles obtained from the recursive rollout.
We form calibrated bounds
\[
\widehat{\bm{q}}_{i,t+h}^{(\ell)\prime}
=
\widehat{\bm{m}}_{i,t+h}
-
s(h)\left(
\widehat{\bm{m}}_{i,t+h}
-
\widehat{\bm{q}}_{i,t+h}^{(\ell)}
\right),
\qquad
\widehat{\bm{q}}_{i,t+h}^{(u)\prime}
=
\widehat{\bm{m}}_{i,t+h}
+
s(h)\left(
\widehat{\bm{q}}_{i,t+h}^{(u)}
-
\widehat{\bm{m}}_{i,t+h}
\right).
\]
The factor $s(h)$ is selected on VAL to match a target coverage level. As in the point-forecast case, fallback to the GLOBAL model is applied when a cluster fails to improve validation pinball loss.

\section{Analysis of the traffic datasets} \label{sec:real_data}

Our primary objective is not to maximize forecasting accuracy per se, but to demonstrate that forecastability-driven specialization (e.g., via clustering) can improve predictive performance by grouping time series with similar forecasting characteristics. Clustering is performed at the level of entire MTS objects rather than individual components: each cluster contains a subset of MTS, and all components within each series are modeled jointly by the forecaster.

\subsection{Evaluation protocol and metrics}
\label{sec:eval_metrics}

All methods are evaluated under the strict TRAIN/VAL/TEST protocol described earlier. Validation data are used exclusively for cluster reassignment, model selection, and fallback decisions, while TEST is used once for final evaluation, ensuring genuine out-of-sample assessment.

For point forecasting, we report mean squared error (MSE) and mean absolute error (MAE). Let $\ell^{(m)}_{i,h}$ denote the TEST loss for method $m$, series $i$, and horizon $h$. The mean loss is
\[
\mu^{(m)}_h
=
\frac{1}{N}\sum_{i=1}^N \ell^{(m)}_{i,h}.
\]
MSE emphasizes large errors, while MAE reflects typical deviations; reporting both provides complementary insights.

For probabilistic forecasting, we report pinball loss (Equation \eqref{eq:pinball_multi}), averaged over quantiles and components, together with the MSE of the median forecast ($q=0.5$) for comparability with point forecasting.

Although our method uses Huber loss on validation data for reassignment and model selection, all methods are evaluated using the same TEST metrics to ensure a fair comparison.

\textit{Relative gains with respect to GLOBAL}. To quantify improvements, we report relative percentage changes with respect to the GLOBAL baseline, computed using MSE. For a given horizon $h$, the relative gain of method $m$ is
\[
\%\Delta^{(m)}_h
=
100 \cdot
\frac{\mu^{(\mathrm{GLOBAL})}_h - \mu^{(m)}_h}{\mu^{(\mathrm{GLOBAL})}_h}.
\]
A positive value indicates reduced forecasting error relative to GLOBAL, while a negative value indicates degradation.

\textit{Series-level benefit and fallback diagnostics}. Mean losses may conceal heterogeneous effects across series. We therefore report two additional series-level quantities, both computed using MSE.

The \textbf{benefit fraction} at horizon $h$ measures the proportion of series for which method $m$ improves upon GLOBAL on TEST data:
\[
\mathrm{Ben}^{(m)}_h
=
\frac{1}{N}\sum_{i=1}^N
\mathbf{1}\!\left\{
\ell^{(m)}_{i,h}
<
\ell^{(\mathrm{GLOBAL})}_{i,h}
\right\}.
\]
In the tables, we report $\mathrm{Ben}^{(m)}_h$ as a percentage (Ben\%).

The \textbf{fallback fraction} quantifies how often specialization is rejected based on validation data. Fallback decisions are made at the cluster level: let $\mathcal{K}_{\mathrm{fb}}$ denote the set of clusters flagged as non-specializable on VAL, i.e., those for which the average validation loss of the specialized prototype exceeds that of the GLOBAL model on the same set of series. Fallback is then applied at the series level: any series assigned to a flagged cluster is routed to the GLOBAL forecaster at TEST time. The fallback fraction is therefore
\[
\mathrm{Fb}
=
\frac{1}{N}\sum_{i=1}^N
\mathbf{1}\!\left\{
c_i \in \mathcal{K}_{\mathrm{fb}}
\right\}.
\]
In the tables, we report $\mathrm{Fb}$ as a percentage (Fb\%). A low fallback fraction indicates stable specialization, while higher values reflect active protection against negative transfer.

Taken together, the relative gains, benefit fraction, and fallback fraction characterize not only the magnitude of improvement, but also its breadth and robustness across series.

\subsection{Comparison methods and motivation}
\label{sec:com_methods}

We consider the following baselines, each isolating a distinct aspect of specialization:
\begin{itemize} 
\item \textbf{GLOBAL}: a single pooled forecaster trained on all series.
\item \textbf{INDIVIDUAL}: one forecaster per series, representing full specialization.
\item \textbf{{FEAT-KMEANS}}: clustering based on simple training-set summaries (means and standard deviations), followed by one prototype per cluster. This captures coarse heterogeneity while excluding temporal dynamics and forecasting loss.
\item \textbf{{RANDOM-BALANCED}}: balanced random clustering, isolating the effect of parameter sharing without data-driven grouping.
\item \textbf{{CLUSTER} (OURS)}: clustering driven by validation forecasting loss, aligning specialization with forecastability.
\end{itemize}
For all cluster-based methods, specialization is safeguarded by a validation-based fallback mechanism to mitigate negative transfer. For each candidate number of clusters $K$, we evaluate five random initializations and select the best-performing configuration on the validation set using a method-specific criterion (Huber loss for our method, MSE for the baselines).

\subsection{The PEMS-BAY data}

The PEMS-BAY dataset contains traffic speed measurements collected from 325 loop detectors from the PEMS deployed on major highways in the San Francisco Bay Area, California. The measurements are recorded every 5 minutes \citep{li2018dcrnn_traffic}, so each complete day contains 288 observations. We retain only days with full 24-hour recordings, that is, days with exactly 288 time points, and discard incomplete days to obtain a fixed-length representation.

After preprocessing, the dataset contains \(N=180\) MTS, each of dimension \(T \times P = 288 \times 325\). Each MTS comprises one day of observations recorded at 5-minute intervals from \(P=325\) spatially distributed traffic sensors, yielding \(T=288\) time points per day. Thus, each observational unit is a multivariate time series spanning a single day, whereas the individual sensors constitute its components. This structure makes the dataset well suited for evaluating high-dimensional MTS forecasting methods under strong temporal and spatial dependence. The dataset is publicly available online\footnote{\url{https://zenodo.org/records/5724362}}.

\subsubsection{Point forecasting for the PEMS-BAY data}

Table~\ref{tab:pemsbay_global_indiv} reports the TEST forecasting accuracy of the GLOBAL and INDIVIDUAL baselines. Across all horizons, the GLOBAL model consistently outperforms the INDIVIDUAL model in both MSE and MAE. This indicates that pooling information across MTS is more effective than fitting separate models to each series, likely because the GLOBAL model exploits a larger effective sample and better captures shared temporal and cross-sensor dependence. At the same time, this result highlights the trade-off between information sharing and cross-series heterogeneity, motivating the proposed forecastability-driven clustering framework.

\begin{table}[htbp]
\centering
\caption{TEST forecasting accuracy ($\times 100$) for the GLOBAL and INDIVIDUAL baselines on PEMS-BAY.}
\label{tab:pemsbay_global_indiv}
\begin{tabular}{lcccccc}
\toprule
Method 
& \multicolumn{2}{c}{$h=1$}
& \multicolumn{2}{c}{$h=3$}
& \multicolumn{2}{c}{$h=6$} \\
\cmidrule(lr){2-3}\cmidrule(lr){4-5}\cmidrule(lr){6-7}
& MSE & MAE & MSE & MAE & MSE & MAE \\
\midrule
GLOBAL      & 7.58 & 15.21 & 7.98 & 16.39 & 9.24 & 19.39 \\
INDIVIDUAL  & 10.11 & 20.86 & 10.64 & 21.77 & 11.56 & 23.26 \\
\bottomrule
\end{tabular}
\end{table}

Table~\ref{tab:pemsbay_cluster_test_h136} reports the TEST performance of the clustering-based approaches across different numbers of clusters \(K\) and horizons \(h \in \{1,3,6\}\), corresponding to 5-, 15-, and 30-minute-ahead forecasts. All TEST metrics are computed \emph{after} applying the validation-based fallback safeguard: any series assigned to a cluster identified as non-specializable on VAL is routed to the GLOBAL forecaster at evaluation time. We also report the fallback rate (Fb\%) and benefit rate (Ben\%), defined in Section~\ref{sec:eval_metrics}.

Overall, the proposed VAL-driven method delivers the strongest and most consistent improvements over GLOBAL across all horizons. The best-performing configuration is attained at \(K^\star=7\), where TEST MSE reductions reach $18.12\%$, $19.24\%$, and $23.75\%$ for $h=1,3,6$, respectively. The corresponding MAE reductions follow the same pattern, indicating that the gains are stable across error measures. Compared with \textsc{Feat-Kmeans} and \textsc{Random-Balanced}, the proposed method also achieves substantially higher benefit rates and essentially zero fallback. This indicates that the validation-Huber-based reassignment identifies clusters that are genuinely forecast-specializable, rather than requiring frequent fallback to the GLOBAL model. Additional details on the selection of \(K\) are provided in the Supplementary Material (Tables~\ref{tab:pemsbay_val_select_feat}--\ref{tab:pemsbay_val_select_ours}).

\begin{table}[!htbp]
\centering
\caption{PEMS-BAY TEST results (scaled by $\times 100$) for horizons $h\in\{1,3,6\}$. Bold indicates the selected $K^\star$ for each method.}
\label{tab:pemsbay_cluster_test_h136}
\small
\resizebox{\linewidth}{!}{%
\begin{tabular}{cc rrrrrr rrrrrr rrrrrr}
\toprule
\multirow{2}{*}{$K$} & \multirow{2}{*}{$h$}
& \multicolumn{6}{c}{\textsc{Feat-Kmeans} (VAL MSE)}
& \multicolumn{6}{c}{\textsc{Random-Balanced} (VAL MSE)}
& \multicolumn{6}{c}{\textsc{CLUSTER (Ours)} (VAL Huber)} \\
\cmidrule(lr){3-8}\cmidrule(lr){9-14}\cmidrule(lr){15-20}
& & MSE & $\Delta$\% & MAE & $\Delta$\% & Ben\% & Fb\%
  & MSE & $\Delta$\% & MAE & $\Delta$\% & Ben\% & Fb\%
  & MSE & $\Delta$\% & MAE & $\Delta$\% & Ben\% & Fb\% \\
\midrule

\multirow{3}{*}{2}
& 1 & 7.43 & +1.92 & 14.81 & +2.60 & 27 & \multirow{3}{*}{71}
    & 7.39 & +2.47 & 15.09 & +0.76 & 34 & \multirow{3}{*}{0}
    & 7.00 & +7.63 & 14.70 & +3.36 & 36 & \multirow{3}{*}{0} \\
& 3 & 7.73 & +3.18 & 15.65 & +4.55 & 29 &
    & 7.66 & +4.03 & 15.76 & +3.85 & 62 &
    & 7.19 & +9.88 & 15.11 & +7.85 & 71 &  \\
& 6 & 8.65 & +6.36 & 17.85 & +7.94 & 29 &
    & 8.49 & +8.13 & 17.56 & +9.44 & 80 &
    & 7.68 & +16.91 & 16.16 & +16.65 & 91 &  \\
\midrule

\multirow{3}{*}{3}
& 1 & 7.44 & +1.80 & 14.83 & +2.48 & 27 & \multirow{3}{*}{71}
    & 7.13 & +5.83 & 14.96 & +1.66 & 36 & \multirow{3}{*}{0}
    & 6.63 & +12.48 & 14.00 & +7.97 & 78 & \multirow{3}{*}{0} \\
& 3 & 7.75 & +2.96 & 15.69 & +4.27 & 28 &
    & 7.46 & +6.54 & 15.80 & +3.62 & 53 &
    & 6.91 & +13.43 & 14.73 & +10.14 & 84 &  \\
& 6 & 8.69 & +6.00 & 17.93 & +7.49 & 29 &
    & 8.45 & +8.57 & 18.11 & +6.59 & 61 &
    & 7.64 & +17.32 & 16.52 & +14.78 & 91 &  \\
\midrule

\multirow{3}{*}{4}
& 1 & 7.31 & +3.50 & 14.74 & +3.08 & 42 & \multirow{3}{*}{49}
    & 7.07 & +6.74 & 15.01 & +1.33 & 28 & \multirow{3}{*}{25}
    & 6.63 & +12.46 & 14.05 & +7.66 & 77 & \multirow{3}{*}{0} \\
& 3 & 7.65 & +4.22 & 15.62 & +4.70 & 43 &
    & 7.50 & +6.00 & 16.18 & +1.29 & 31 &
    & 6.94 & +13.04 & 14.80 & +9.68 & 79 &  \\
& 6 & 8.61 & +6.79 & 17.78 & +8.30 & 44 &
    & 9.08 & +1.74 & 19.67 & -1.48 & 26 &
    & 7.89 & +14.65 & 17.03 & +12.15 & 75 &  \\
\midrule

\multirow{3}{*}{5}
& 1 & 6.87 & +9.30 & 14.51 & +4.63 & 41 & \multirow{3}{*}{45}
    & \textbf{7.05} & \textbf{+6.97} & \textbf{15.01} & \textbf{+1.34} & \textbf{32} & \multirow{3}{*}{\textbf{20}}
    & 6.52 & +13.97 & 14.13 & +7.07 & 70 & \multirow{3}{*}{0} \\
& 3 & 7.18 & +10.05 & 15.35 & +6.37 & 43 &
    & \textbf{7.39} & \textbf{+7.41} & \textbf{15.88} & \textbf{+3.11} & \textbf{42} &
    & 6.82 & +14.59 & 14.90 & +9.09 & 73 &  \\
& 6 & 8.18 & +11.50 & 17.74 & +8.49 & 39 &
    & \textbf{8.53} & \textbf{+7.69} & \textbf{18.36} & \textbf{+5.27} & \textbf{50} &
    & 7.82 & +15.39 & 17.32 & +10.66 & 73 &  \\
\midrule

\multirow{3}{*}{6}
& 1 & 6.82 & +10.01 & 14.38 & +5.44 & 48 & \multirow{3}{*}{42}
    & 7.04 & +7.08 & 14.96 & +1.62 & 36 & \multirow{3}{*}{17}
    & 6.38 & +15.84 & 13.80 & +9.28 & 89 & \multirow{3}{*}{0} \\
& 3 & 7.13 & +10.63 & 15.21 & +7.22 & 47 &
    & 7.43 & +6.98 & 15.98 & +2.51 & 45 &
    & 6.73 & +15.63 & 14.74 & +10.07 & 88 &  \\
& 6 & 8.01 & +13.35 & 17.27 & +10.94 & 47 &
    & 8.80 & +4.83 & 18.89 & +2.58 & 47 &
    & 7.80 & +15.64 & 17.18 & +11.36 & 78 &  \\
\midrule

\multirow{3}{*}{7}
& 1 & 6.77 & +10.61 & 14.34 & +5.70 & 51 & \multirow{3}{*}{37}
    & 6.95 & +8.26 & 14.82 & +2.55 & 36 & \multirow{3}{*}{29}
    & \textbf{6.20} & \textbf{+18.12} & \textbf{13.46} & \textbf{+11.51} & \textbf{84} & \multirow{3}{*}{\textbf{0}} \\
& 3 & 7.06 & +11.53 & 15.08 & +7.99 & 51 &
    & 7.26 & +9.08 & 15.59 & +4.89 & 51 &
    & \textbf{6.45} & \textbf{+19.24} & \textbf{14.01} & \textbf{+14.54} & \textbf{86} &  \\
& 6 & 7.86 & +14.98 & 16.97 & +12.45 & 53 &
    & 8.22 & +11.09 & 17.70 & +8.68 & 59 &
    & \textbf{7.05} & \textbf{+23.75} & \textbf{15.48} & \textbf{+20.14} & \textbf{92} &  \\
\midrule

\multirow{3}{*}{8}
& 1 & 6.61 & +12.72 & 14.13 & +7.11 & 39 & \multirow{3}{*}{59}
    & 6.98 & +7.85 & 14.90 & +2.07 & 27 & \multirow{3}{*}{37}
    & 6.36 & +16.03 & 13.70 & +9.94 & 74 & \multirow{3}{*}{0} \\
& 3 & 6.91 & +13.40 & 14.98 & +8.59 & 40 &
    & 7.31 & +8.48 & 15.74 & +3.99 & 41 &
    & 6.64 & +16.85 & 14.42 & +12.05 & 78 &  \\
& 6 & 7.76 & +16.01 & 17.06 & +12.00 & 40 &
    & 8.27 & +10.51 & 17.93 & +7.50 & 51 &
    & 7.43 & +19.57 & 16.29 & +15.97 & 83 &  \\
\midrule

\multirow{3}{*}{9}
& 1 & \textbf{6.59} & \textbf{+13.05} & \textbf{14.08} & \textbf{+7.42} & \textbf{48} & \multirow{3}{*}{\textbf{39}}
    & 6.91 & +8.74 & 14.71 & +3.27 & 37 & \multirow{3}{*}{33}
    & 6.43 & +15.17 & 13.73 & +9.72 & 77 & \multirow{3}{*}{6} \\
& 3 & \textbf{6.84} & \textbf{+14.25} & \textbf{14.77} & \textbf{+9.87} & \textbf{54} &
    & 7.22 & +9.50 & 15.52 & +5.29 & 47 &
    & 6.76 & +15.35 & 14.57 & +11.11 & 76 &  \\
& 6 & \textbf{7.53} & \textbf{+18.54} & \textbf{16.48} & \textbf{+14.98} & \textbf{57} &
    & 8.25 & +10.68 & 17.92 & +7.54 & 49 &
    & 7.91 & +14.40 & 17.01 & +12.26 & 70 &  \\
\bottomrule
\end{tabular}}
\end{table}

\subsubsection{Probabilistic forecasting for the PEMS-BAY data} \label{prob_fore_bay}

For probabilistic forecasting, we estimate the conditional quantiles
$q \in \{0.1,0.5,0.9\}$. The median quantile ($q=0.5$) is used as the point forecast, while the lower and upper quantiles ($q=0.1$ and $q=0.9$) define a central 80\% prediction interval. Forecast quality is evaluated by the pinball loss, the MSE of the median forecast, and the coverage and width of the resulting prediction intervals. The INDIVIDUAL baseline is not reported because its forecasting performance is substantially worse than that of GLOBAL and all clustering-based methods.

Based on the validation pinball loss, the selected numbers of clusters are $K^*=7$ for \textsc{Feat-Kmeans}, $K^*=5$ for \textsc{Random-Balanced}, and $K^*=5$ for our proposed method.

\begin{figure}[ht]
    \centering
    \includegraphics[width=.8\linewidth]{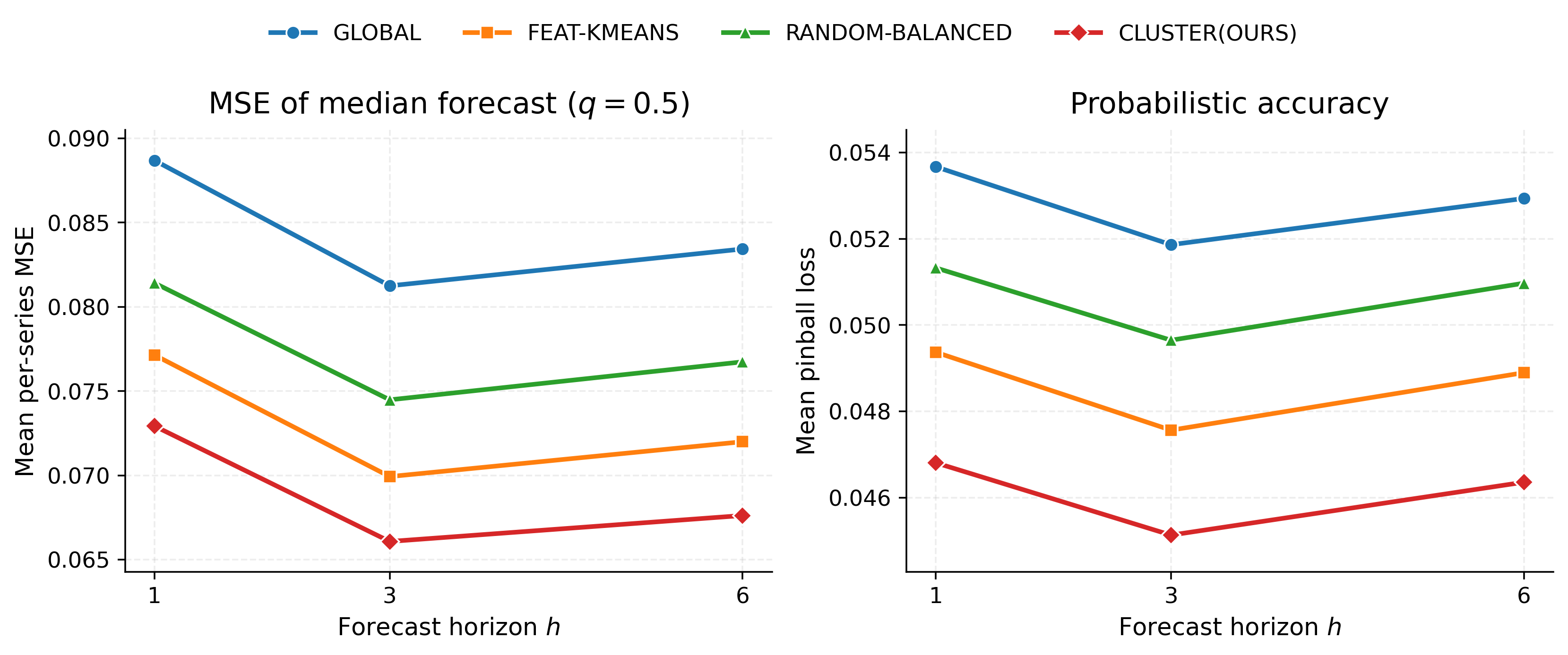}
    \caption{Probabilistic forecasting accuracy across prediction horizons $h \in \{1,3,6\}$ using the selected $K^\star$ for the PEMS-BAY data.}
    \label{fig:pinball_median}
\end{figure}

Figure~\ref{fig:pinball_median} compares TEST forecasting accuracy across horizons using the selected $K^\star$. The left panel reports the mean per-series MSE of the median forecast, while the right panel reports the mean pinball loss. Across all horizons, the proposed method attains the lowest MSE and pinball loss, showing that it improves both point and probabilistic forecasting relative to GLOBAL and the baseline clustering strategies.

\begin{figure}[ht]
    \centering
    \includegraphics[width=.8\linewidth]{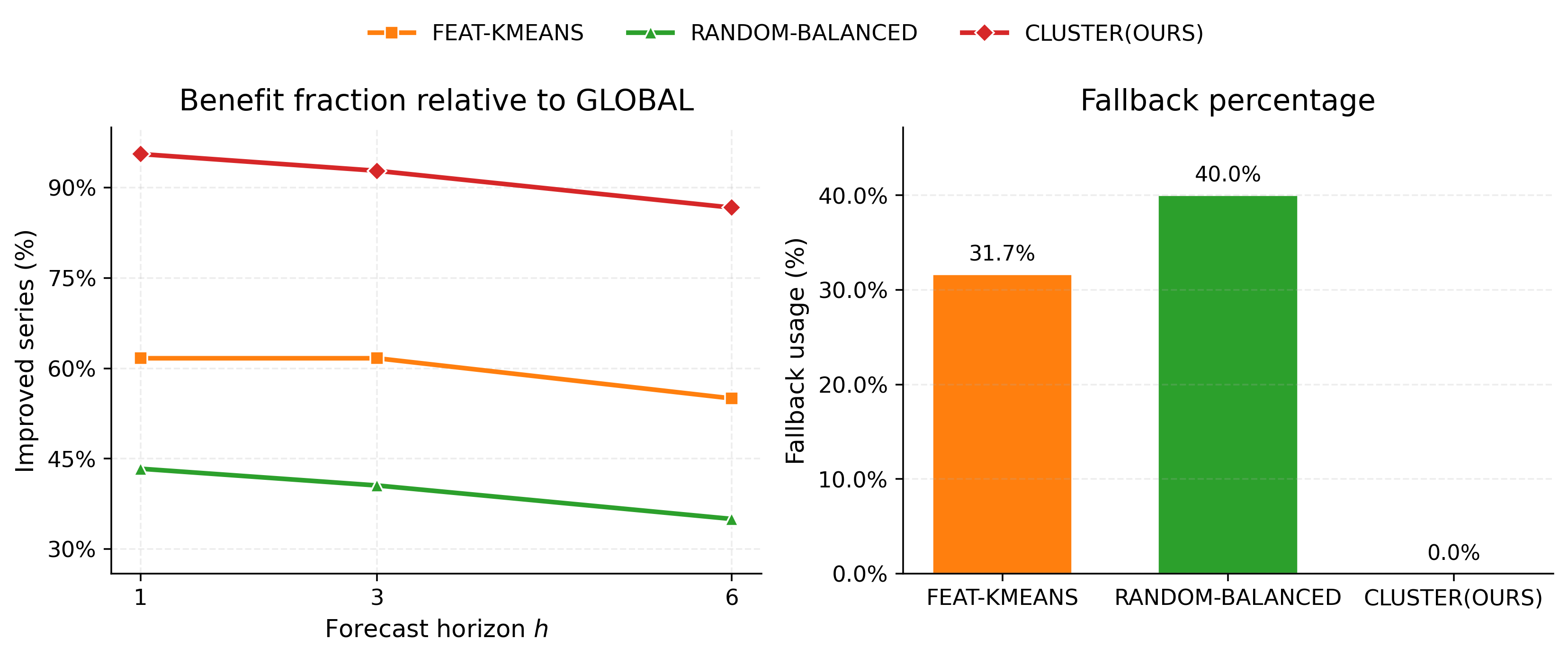}
    \caption{Practical impact of clustering-based forecasting in terms of benefit fraction and fallback usage for the PEMS-BAY data.}
    \label{fig:ben_fall}
\end{figure}

Figure~\ref{fig:ben_fall} highlights the practical impact of clustering. The proposed method improves performance for a large proportion of series across all horizons, while requiring much less fallback than the baseline clustering approaches.

\begin{figure}[ht]
    \centering
    \includegraphics[width=.8\linewidth]{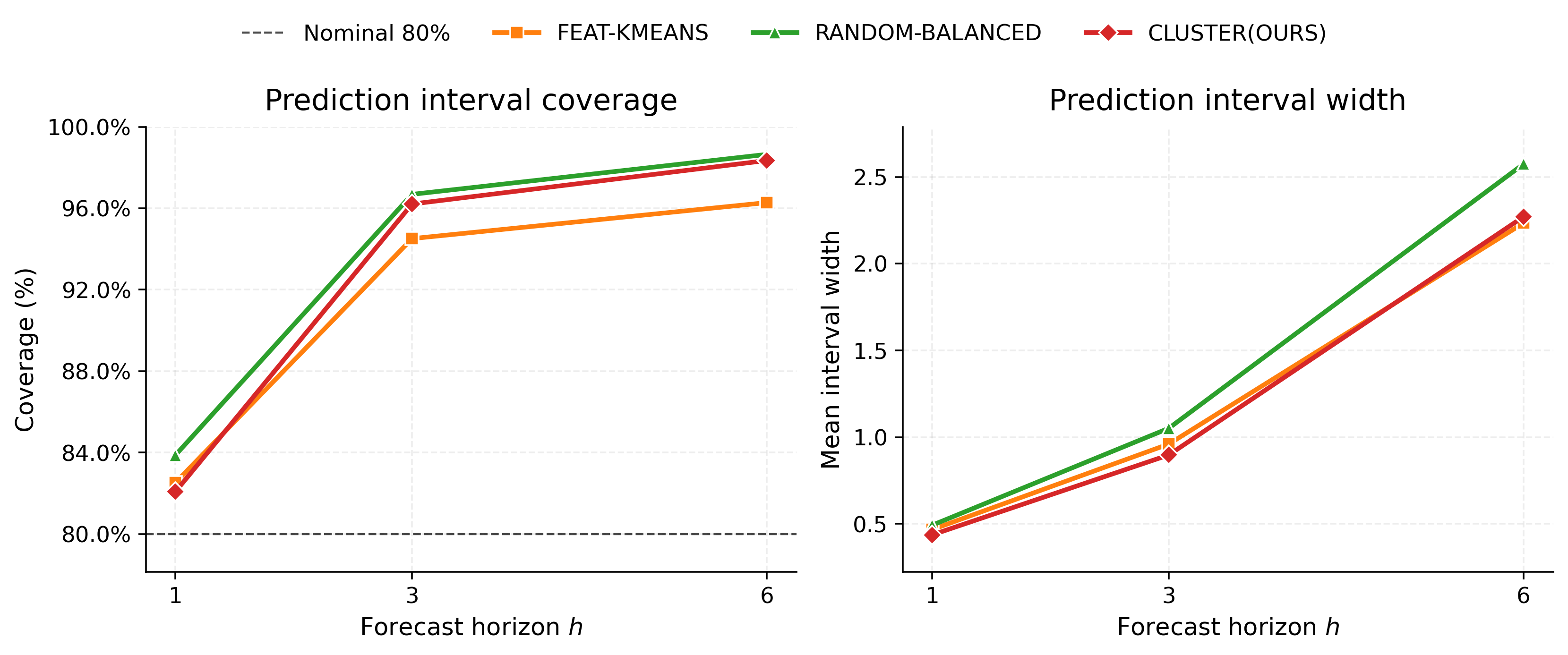}
    \caption{Calibration and sharpness of probabilistic forecasts for the PEMS-BAY data.}
    \label{fig:cal_sha}
\end{figure}

Figure~\ref{fig:cal_sha} examines calibration and sharpness. The proposed approach achieves strong coverage with competitive interval widths, indicating reliable and effective uncertainty quantification.

\subsection{The PEMS-SF data}

The PEMS-SF dataset is a high-dimensional MTS benchmark derived from the California Performance Measurement System (PEMS). It records freeway lane occupancy in the San Francisco Bay Area at 10-minute intervals from January 1, 2008, to March 30, 2009. Following standard preprocessing—removal of public holidays and days affected by daylight saving time—the dataset comprises $N=440$ daily MTS objects. Each object corresponds to a single day with $T=144$ time points (00{:}00--23{:}50) and $P=963$ sensors, yielding a $144 \times 963$ time series. The dataset is publicly available from the UCI Machine Learning Repository\footnote{\url{https://archive.ics.uci.edu/dataset/204/pems+sf}}.

Forecasting performance is again evaluated at horizons $h \in \{1,3,6\}$, corresponding to 10-, 30-, and 60-minute-ahead predictions, respectively, allowing assessment across short- and moderate-term forecasting regimes.

\subsubsection{Point forecasting for the PEMS-SF data}

Table~\ref{tab:pems_global_indiv} reports TEST forecasting accuracy for the two non-clustering baselines on PEMS-SF. GLOBAL consistently outperforms INDIVIDUAL in both MSE and MAE across all horizons, reflecting the benefits of pooling in short, high-dimensional series. Errors increase with the forecasting horizon, as expected in traffic prediction. Selection of $K$ is based on validation performance; detailed results across random initializations are provided in the Supplementary Material (Tables~\ref{tab:val_select_featkmeans}--\ref{tab:val_select_ours}).

\begin{table}[h]
\centering
\footnotesize
\caption{TEST forecasting accuracy ($\times 100$) for the GLOBAL and INDIVIDUAL baselines on PEMS-SF.}
\label{tab:pems_global_indiv}
\setlength{\tabcolsep}{6pt}
\renewcommand{\arraystretch}{1.15}
\begin{tabular}{lcccccc}
\toprule
\multirow{2}{*}{Method}
& \multicolumn{2}{c}{$h=1$} & \multicolumn{2}{c}{$h=3$} & \multicolumn{2}{c}{$h=6$} \\
\cmidrule(lr){2-3}\cmidrule(lr){4-5}\cmidrule(lr){6-7}
& MSE & MAE & MSE & MAE & MSE & MAE \\
\midrule
GLOBAL
& 12.45 & 17.74
& 13.91 & 20.67
& 20.52 & 30.00 \\
INDIVIDUAL
& 29.61 & 40.34
& 31.08 & 41.75
& 33.77 & 43.70 \\
\bottomrule
\end{tabular}
\end{table}

Overall, validation loss decreases with $K$, indicating that moderate clustering captures predictive heterogeneity, but gains plateau beyond $K=7$. The penalized criterion therefore selects $K^\star=7$ as the optimal trade-off between accuracy and complexity.

Table~\ref{tab:pems_test_byK_landscape} reports TEST improvements relative to GLOBAL, along with benefited series (Ben\%) and fallback usage (Fb\%). All TEST metrics are computed after applying the validation-based fallback safeguard.

\textsc{Ours} delivers the largest and most stable gains across horizons. At $K^\star=7$, it achieves consistent reductions in MSE and MAE, with broad improvements across series (Ben\% = $98\%$, $97\%$, and $89\%$ for $h=1,3,6$) and zero fallback, indicating reliable forecast-specializable clusters. In contrast, \textsc{Feat-Kmeans} and \textsc{Random-Balanced} exhibit substantially higher fallback rates across many values of $K$, suggesting that their clusters are often not forecast-specializable. The fallback mechanism is therefore essential to prevent harmful specialization under unstable assignments.

\begin{table}[t]
\centering
\small
\caption{PEMS-SF TEST results (scaled by $\times 100$) for horizons $h\in\{1,3,6\}$. Bold indicates the selected $K^\star$ for each method.}
\label{tab:pems_test_byK_landscape}
\setlength{\tabcolsep}{3.2pt}
\renewcommand{\arraystretch}{1.08}
\resizebox{\linewidth}{!}{%
\begin{tabular}{cc
  r r r r r c
  r r r r r c
  r r r r r c}
\toprule
\multirow{2}{*}{$K$} & \multirow{2}{*}{$h$}
& \multicolumn{6}{c}{\textsc{Feat-Kmeans} (VAL MSE)}
& \multicolumn{6}{c}{\textsc{Random-Balanced} (VAL MSE)}
& \multicolumn{6}{c}{\textsc{CLUSTER (Ours)} (VAL Huber)} \\
\cmidrule(lr){3-8}\cmidrule(lr){9-14}\cmidrule(lr){15-20}
&& MSE & $\Delta$\% & MAE & $\Delta$\% & Ben\% & Fb\%
& MSE & $\Delta$\% & MAE & $\Delta$\% & Ben\% & Fb\%
& MSE & $\Delta$\% & MAE & $\Delta$\% & Ben\% & Fb\% \\
\midrule

\multirow{3}{*}{2} & 1
& 11.62 & +6.65 & 16.00 & +9.77 & 86 & \multirow{3}{*}{0}
& 12.56 & -0.94 & 18.35 & -3.45 & 13 & \multirow{3}{*}{50}
& 11.68 & +6.18 & 16.01 & +9.72 & 28 & \multirow{3}{*}{72} \\
& 3
& 12.96 & +6.84 & 18.97 & +8.22 & 62 &
& 15.09 & -8.42 & 23.18 & -12.11 & 6 &
& 12.64 & +9.18 & 18.08 & +12.53 & 28 & \\
& 6
& 18.67 & +9.01 & 27.76 & +7.46 & 56 &
& 24.72 & -20.47 & 35.70 & -18.99 & 1 &
& 17.16 & +16.39 & 25.39 & +15.38 & 28 & \\
\midrule

\multirow{3}{*}{3} & 1
& 11.83 & +4.97 & 16.39 & +7.58 & 47 & \multirow{3}{*}{51}
& 12.53 & -0.64 & 18.10 & -2.06 & 7 & \multirow{3}{*}{67}
& 11.37 & +8.65 & 15.77 & +11.08 & 90 & \multirow{3}{*}{0} \\
& 3
& 13.24 & +4.82 & 19.37 & +6.28 & 36 &
& 14.54 & -4.52 & 22.05 & -6.66 & 3 &
& 13.53 & +2.79 & 20.25 & +2.03 & 73 & \\
& 6
& 19.51 & +4.93 & 28.63 & +4.58 & 30 &
& 22.53 & -9.77 & 32.89 & -9.65 & 3 &
& 22.92 & -11.69 & 32.38 & -7.95 & 46 & \\
\midrule

\multirow{3}{*}{4} & 1
& 12.01 & +3.52 & 16.61 & +6.35 & 48 & \multirow{3}{*}{39}
& 12.50 & -0.42 & 18.17 & -2.46 & 15 & \multirow{3}{*}{50}
& 11.20 & +10.02 & 14.99 & +15.45 & 86 & \multirow{3}{*}{7} \\
& 3
& 13.41 & +3.65 & 19.42 & +6.07 & 44 &
& 14.89 & -7.04 & 22.79 & -10.23 & 10 &
& 12.46 & +10.42 & 17.76 & +14.08 & 85 & \\
& 6
& 19.26 & +6.16 & 27.87 & +7.11 & 38 &
& 24.22 & -17.99 & 34.97 & -16.56 & 10 &
& 18.26 & +11.04 & 26.52 & +11.59 & 77 & \\
\midrule

\multirow{3}{*}{5} & 1
& 12.16 & +2.35 & 16.78 & +5.38 & 49 & \multirow{3}{*}{26}
& 12.23 & +1.74 & 17.50 & +1.35 & 32 & \multirow{3}{*}{40}
& 10.65 & +14.44 & 13.52 & +23.78 & 97 & \multirow{3}{*}{0} \\
& 3
& 13.92 & -0.03 & 20.19 & +2.35 & 40 &
& 13.91 & +0.04 & 20.95 & -1.33 & 22 &
& 11.76 & +15.47 & 15.97 & +22.75 & 94 & \\
& 6
& 20.07 & +2.21 & 28.56 & +4.79 & 48 &
& 21.14 & -3.01 & 31.42 & -4.73 & 23 &
& 16.23 & +20.94 & 22.93 & +23.57 & 82 & \\
\midrule

\multirow{3}{*}{\textbf{6}} & 1
& \textbf{11.93} & \textbf{+4.19} & \textbf{16.15} & \textbf{+8.95} & \textbf{55} & \multirow{3}{*}{\textbf{26}}
& 12.45 & +0.00 & 17.73 & +0.00 & 0 & \multirow{3}{*}{100}
& 10.37 & +16.71 & 12.80 & +27.83 & 97 & \multirow{3}{*}{0} \\
& 3
& \textbf{13.27} & \textbf{+4.60} & \textbf{18.77} & \textbf{+9.20} & \textbf{50} &
& 13.91 & +0.00 & 20.67 & +0.00 & 0 &
& 11.26 & +19.07 & 14.90 & +27.90 & 96 & \\
& 6
& \textbf{18.26} & \textbf{+11.04} & \textbf{25.89} & \textbf{+13.71} & \textbf{51} &
& 20.52 & +0.00 & 30.00 & +0.00 & 0 &
& 15.41 & +24.91 & 21.77 & +27.44 & 91 & \\
\midrule

\multirow{3}{*}{\textbf{7}} & 1
& 11.50 & +7.61 & 15.30 & +13.71 & 66 & \multirow{3}{*}{26}
& 12.38 & +0.54 & 17.66 & +0.44 & 7 & \multirow{3}{*}{86}
& \textbf{10.39} & \textbf{+16.55} & \textbf{12.83} & \textbf{+27.66} & \textbf{98} & \multirow{3}{*}{\textbf{0}} \\
& 3
& 13.03 & +6.32 & 18.44 & +10.78 & 50 &
& 13.87 & +0.30 & 20.68 & -0.04 & 6 &
& \textbf{11.27} & \textbf{+18.97} & \textbf{14.87} & \textbf{+28.09} & \textbf{97} & \\
& 6
& 19.02 & +7.32 & 26.82 & +10.59 & 46 &
& 20.53 & -0.02 & 30.18 & -0.61 & 6 &
& \textbf{15.40} & \textbf{+24.95} & \textbf{21.59} & \textbf{+28.02} & \textbf{89} & \\
\midrule

\multirow{3}{*}{8} & 1
& 11.40 & +8.41 & 15.05 & +15.15 & 62 & \multirow{3}{*}{26}
& 12.32 & +1.01 & 17.75 & -0.09 & 28 & \multirow{3}{*}{38}
& 10.77 & +13.45 & 13.83 & +22.02 & 88 & \multirow{3}{*}{11} \\
& 3
& 12.74 & +8.43 & 17.89 & +13.44 & 53 &
& 14.12 & -1.48 & 21.39 & -3.45 & 25 &
& 11.95 & +14.12 & 16.45 & +20.41 & 84 & \\
& 6
& 18.31 & +10.79 & 25.85 & +13.83 & 52 &
& 21.44 & -4.47 & 31.72 & -5.74 & 27 &
& 17.59 & +14.29 & 24.77 & +17.45 & 72 & \\
\midrule

\multirow{3}{*}{\textbf{9}} & 1
& 11.32 & +9.04 & 14.55 & +17.95 & 82 & \multirow{3}{*}{0}
& \textbf{12.33} & \textbf{+0.97} & \textbf{17.65} & \textbf{+0.45} & \textbf{25} & \multirow{3}{*}{\textbf{56}}
& 10.72 & +13.91 & 13.52 & +23.78 & 92 & \multirow{3}{*}{0} \\
& 3
& 13.03 & +6.32 & 18.00 & +12.91 & 71 &
& \textbf{14.13} & \textbf{-1.52} & \textbf{21.26} & \textbf{-2.84} & \textbf{15} &
& 12.15 & +12.68 & 16.57 & +19.85 & 81 & \\
& 6
& 19.59 & +4.55 & 26.80 & +10.68 & 71 &
& \textbf{21.68} & \textbf{-5.62} & \textbf{31.77} & \textbf{-5.89} & \textbf{14} &
& 17.84 & +13.08 & 25.12 & +16.25 & 68 & \\
\bottomrule
\end{tabular}}
\end{table}

\begin{figure}[t]
\centering
\includegraphics[width=0.48\linewidth]{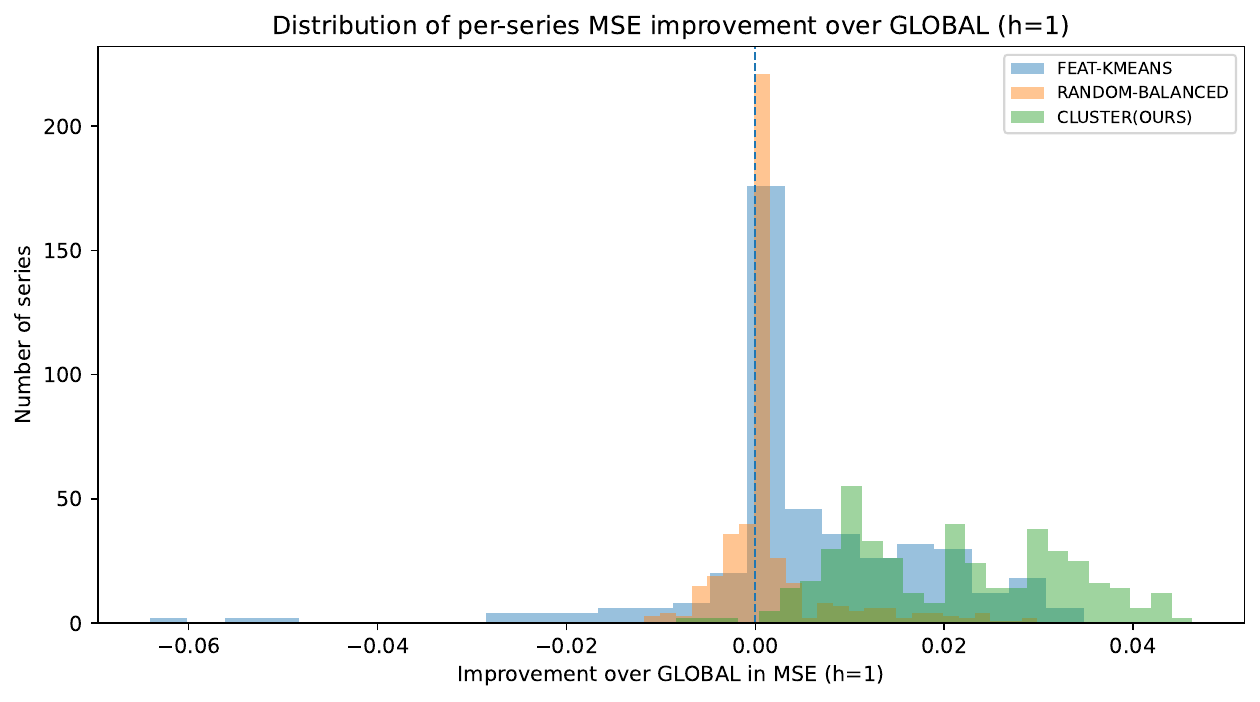}
\includegraphics[width=0.48\linewidth]{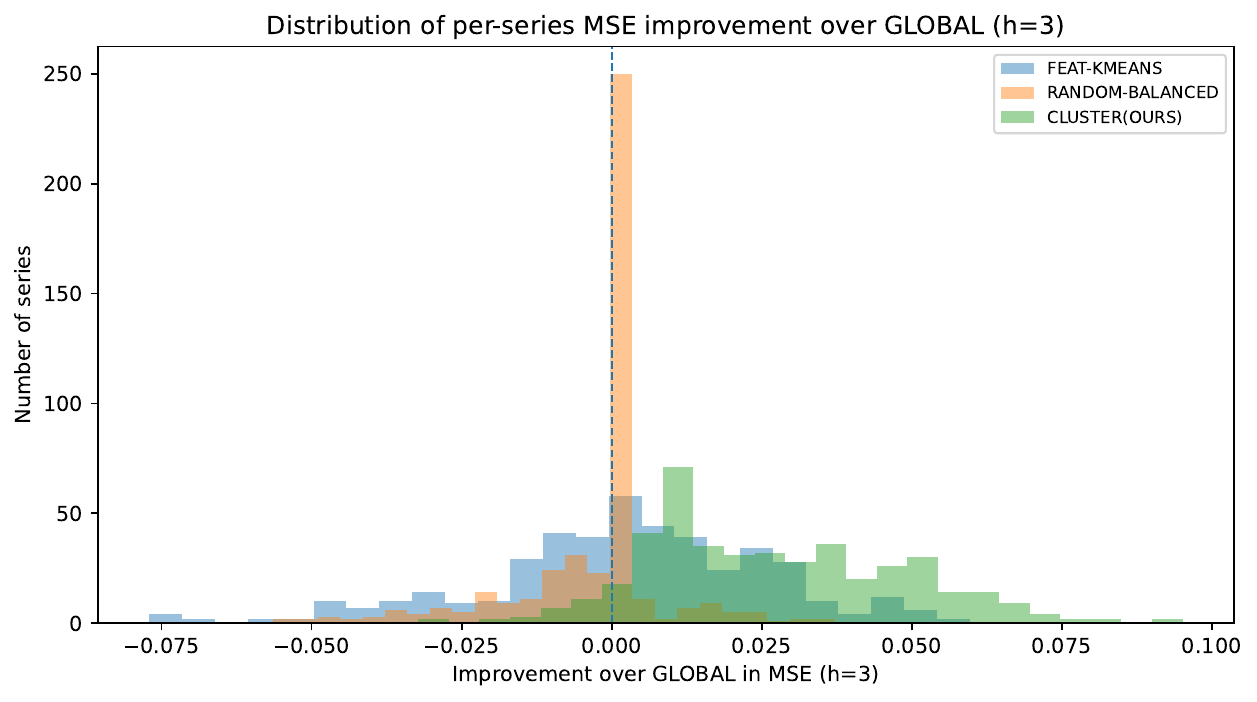}

\vspace{0.2em}
{\small (a) $h=1$ \hspace{3cm} (b) $h=3$}

\vspace{0.4em}

\includegraphics[width=0.5\linewidth]{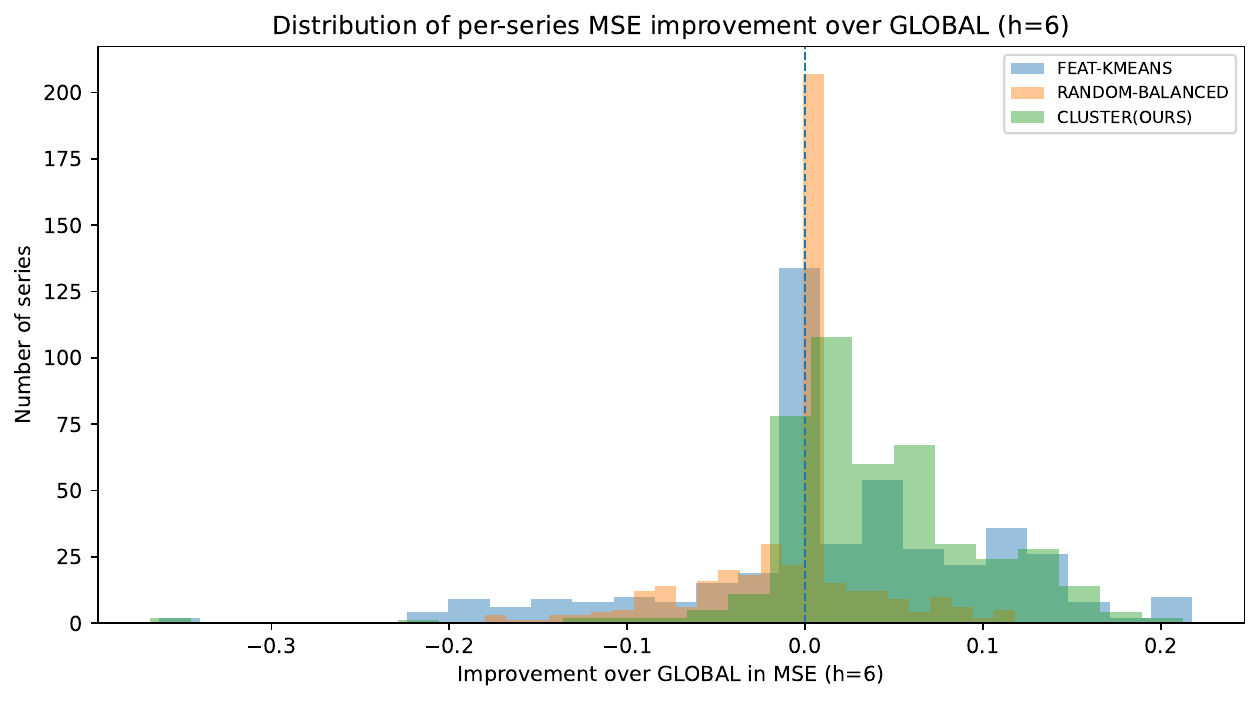}

\vspace{0.2em}
{\small (c) $h=6$}

\caption{Distribution of per-series MSE improvement over GLOBAL at horizons $h=1,3,6$.}
\label{fig:impdist_all}
\end{figure}

Figure~\ref{fig:impdist_all} shows the distribution of per-series MSE improvements over the pooled GLOBAL forecaster. Overall, \textsc{Cluster(Ours)} exhibits the most favorable error profile. At $h=1$ and $h=3$, its distribution is more strongly shifted toward positive values than those of \textsc{Feat-Kmeans} and \textsc{Random-Balanced}, indicating broader and larger gains across series. At $h=6$, the advantage persists but is less pronounced, consistent with the increased difficulty of longer-horizon forecasting.

\begin{figure}[!t]
    \centering
    \includegraphics[width=\linewidth]{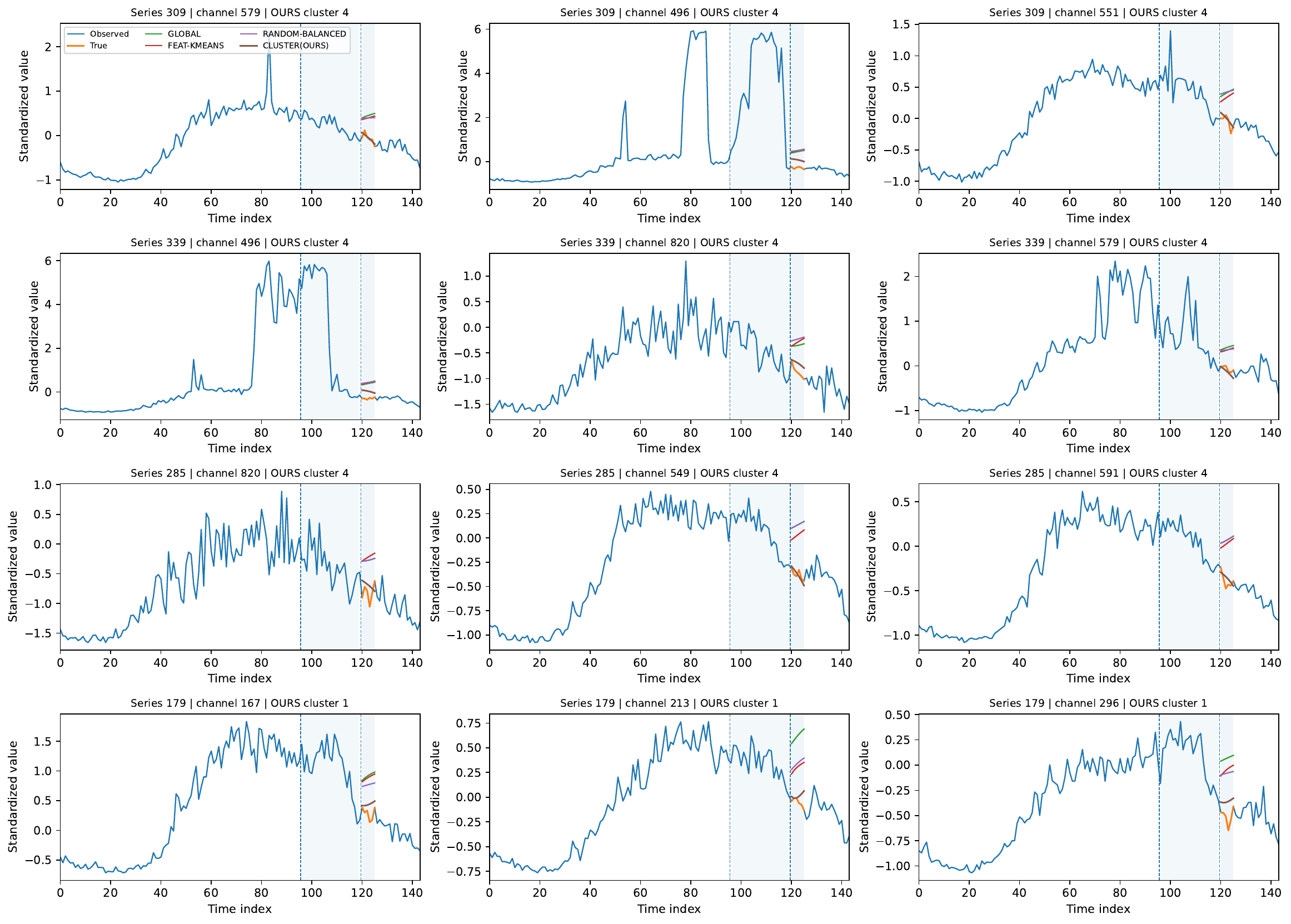}
    \caption{Representative full-day prediction panels at horizon $h=6$. Each subplot compares GLOBAL, \textsc{Feat-Kmeans}, \textsc{Random-Balanced}, and \textsc{Cluster(Ours)} against the true future trajectory for selected series-channel pairs. In these representative examples, \textsc{Cluster(Ours)} tends to remain closer to the held-out future trajectory.}
    \label{fig:rep_panels_h6}
\end{figure}

\begin{figure}[htbp]
    \centering
    \includegraphics[width=\linewidth]{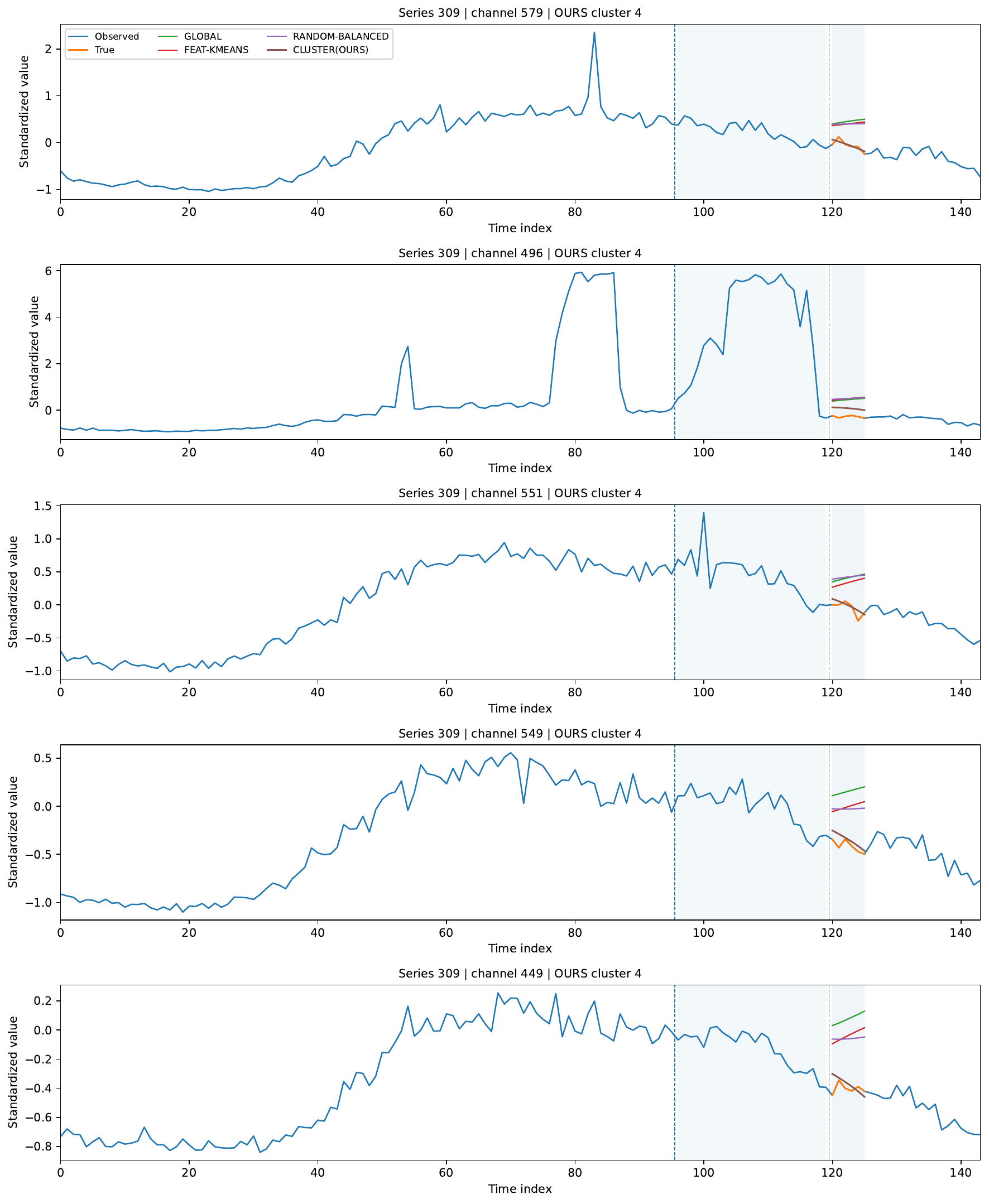}
    \caption{Single-series full-day prediction comparison at horizon \(h=6\) for selected channels of one representative series. The dashed vertical lines mark the VAL and TEST split points. The forecasts from \textsc{Cluster(Ours)} more closely track the held-out future segment in these selected channels, illustrating the trajectory-level improvements reflected in the aggregate error distributions.}
    \label{fig:single_series_h6}
\end{figure}

Figure~\ref{fig:rep_panels_h6} and Figure~\ref{fig:single_series_h6} provide complementary trajectory-level evidence of forecasting gains at horizon $h=6$. The plots display the tail of TRAIN, the VAL segment used for clustering and model selection, and the TEST segment for evaluation. In the representative examples, \textsc{Cluster(Ours)} tracks the realized trajectory more closely than competing methods, particularly during structural changes, where alternatives tend to overshoot or adapt more slowly. The use of the Huber loss further stabilizes forecasts by reducing the influence of large deviations.

Cluster labels (e.g., ``cluster 4'') denote the assigned cluster and corresponding prototype, without implying any ordering. The displayed series and channels are selected to illustrate substantial improvements and are therefore representative rather than random.

In each panel, the solid line denotes observations, dashed vertical lines mark the boundaries of TRAIN/VAL/TEST , and shaded regions indicate the VAL and TEST segments. Forecasts are generated recursively from the last observed window. Additional trajectory-level comparisons at horizon $h=3$ are provided in the Supplementary Material (Figures~\ref{fig:rep_panels_h3} and \ref{fig:single_series_h3}).

\subsubsection{Probabilistic forecasting for the PEMS-SF data} \label{prob_sf_data}

\begin{figure}[ht]
    \centering
    \includegraphics[width=.8\linewidth]{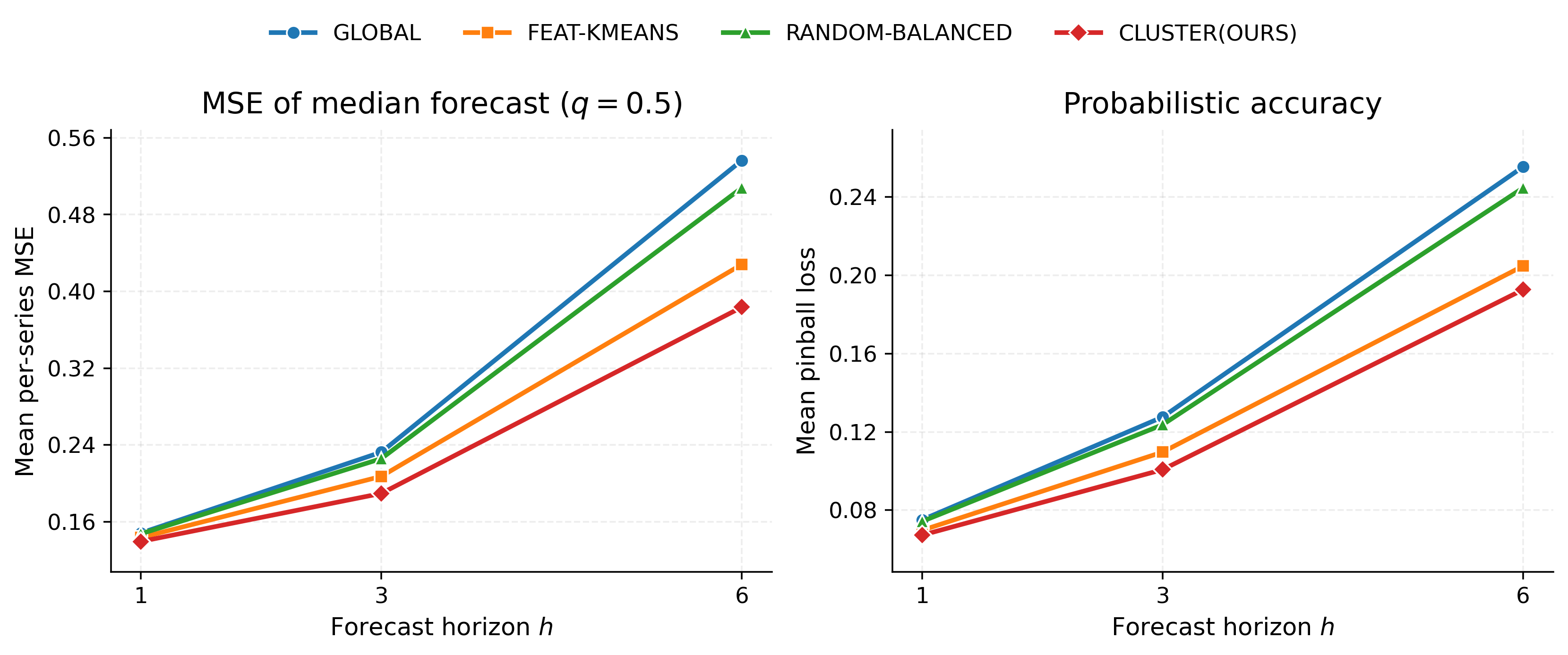}
    \caption{Probabilistic forecasting accuracy across prediction horizons $h \in \{1,3,6\}$ using the selected $K^\star$ for the PEMS-SF data.}
    \label{acc_sf}
\end{figure}

We follow the same setting as in Section~\ref{prob_fore_bay}. Based on the validation pinball loss, the selected numbers of clusters are $K^\star=5$ for \textsc{Feat-Kmeans}, $K^\star=2$ for \textsc{Random-Balanced}, and $K^\star=7$ for \textsc{Cluster(Ours)}. Figure~\ref{acc_sf} shows that \textsc{Cluster(Ours)} achieves the best overall forecasting performance across all horizons. It consistently attains the lowest mean per-series MSE of the median forecast and the lowest mean pinball loss at $h \in \{1,3,6\}$, demonstrating clear improvements over the baseline methods in both point and probabilistic forecasting. 

\begin{figure}[ht]
    \centering
    \includegraphics[width=.8\linewidth]{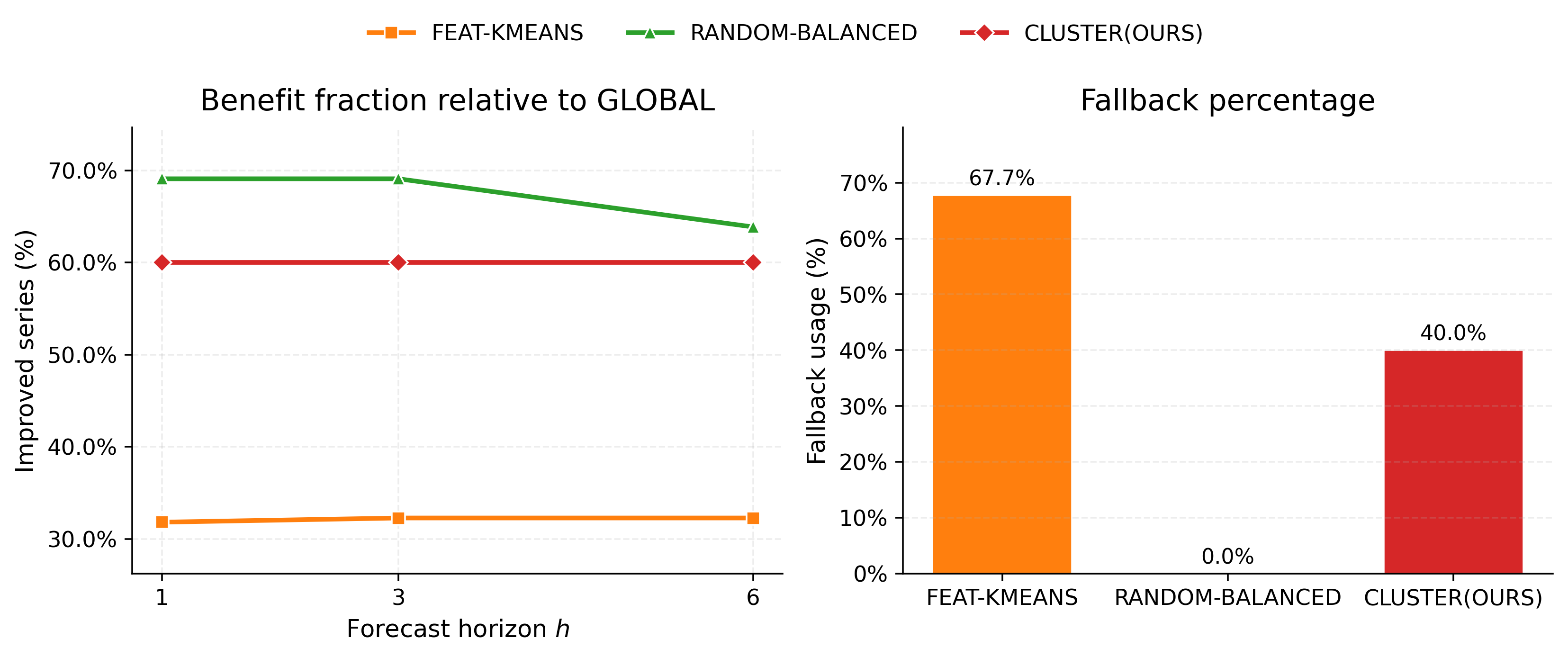}
    \caption{Practical impact of clustering-based forecasting in terms of benefit fraction and fallback usage for the PEMS-SF data.}
    \label{impac_sf}
\end{figure}

Figure~\ref{impac_sf} illustrates the practical effect of clustering. \textsc{Cluster(Ours)} maintains a stable benefit fraction of about $60\%$ across all horizons, while requiring substantially less fallback than \textsc{Feat-Kmeans}. Although \textsc{Random-Balanced} shows a somewhat higher benefit fraction, its aggregate forecasting accuracy remains clearly worse, indicating that benefit fraction alone is not sufficient to characterize overall predictive quality.

\begin{figure}[ht]
    \centering
    \includegraphics[width=.8\linewidth]{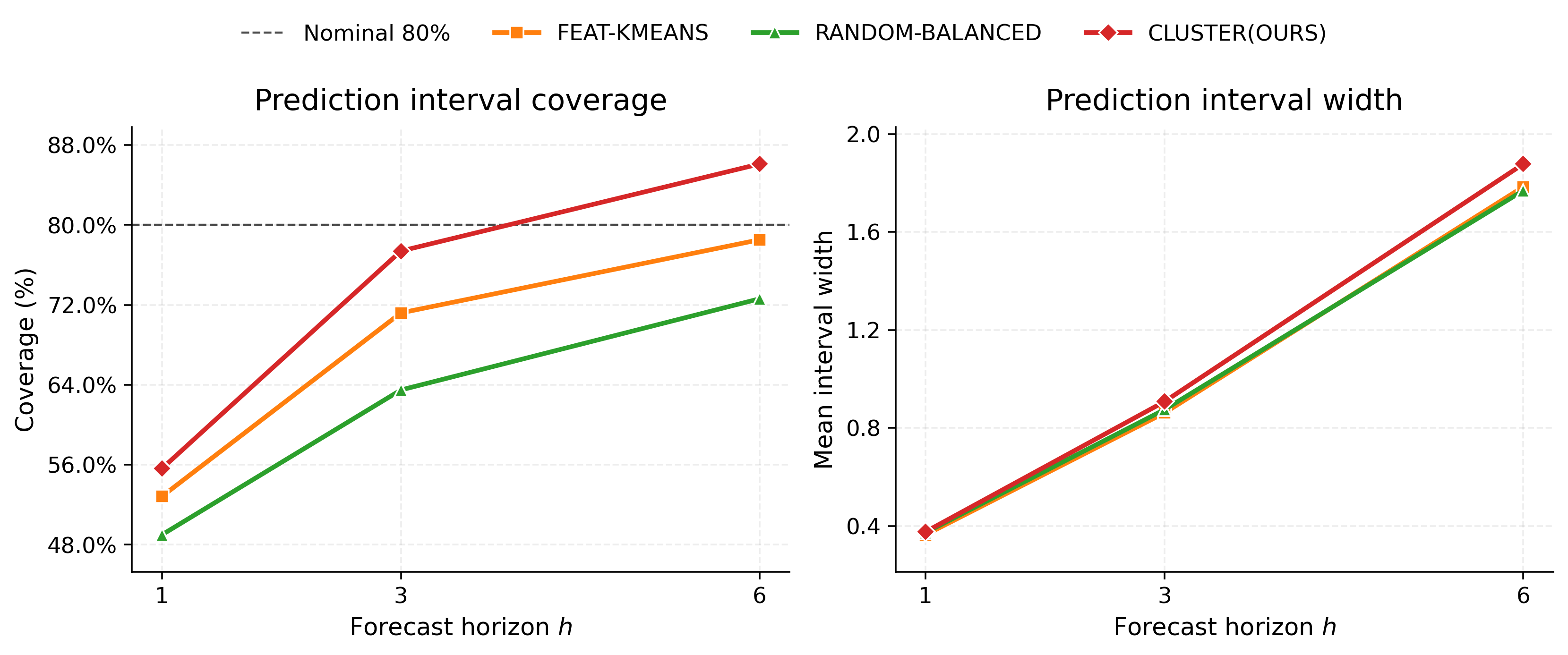}
    \caption{Calibration and sharpness of probabilistic forecasts for the PEMS-SF data.}
    \label{cal_sha_sf}
\end{figure}

The calibration results are shown in Figure~\ref{cal_sha_sf}. \textsc{Cluster(Ours)} achieves the highest prediction interval coverage across all horizons and moves closest to the nominal 80\% level at longer horizons, while incurring only a modest increase in interval width. This indicates a favorable trade-off between calibration and sharpness.

\section{Conclusion} \label{sec:conclusion}

Forecasting high-dimensional MTS is challenging when heterogeneous dynamics render a single pooled model inadequate, while naive specialization can induce unstable assignments and negative transfer. We propose a validation-driven specialization framework in which clusters are defined by out-of-sample predictive performance rather than representation similarity, thereby aligning clustering directly with the forecasting objective.

A key strength of the framework is its strict TRAIN/VAL/TEST protocol, which governs reassignment and model selection in a leakage-free manner. Combined with robust validation losses and a fallback safeguard, the method exploits heterogeneity when beneficial while mitigating negative transfer. This yields a data-driven mechanism for selecting appropriate forecasting models across heterogeneous series and supports adaptive deployment: a newly observed series can be assigned using an initial segment by minimizing short-horizon predictive loss, after which the selected model is used for forecasting. In contrast to prior predictive clustering approaches, which focus primarily on univariate settings and relatively simple linear models, our framework accommodates flexible nonlinear forecasters and scales to high-dimensional MTS. This yields a data-driven mechanism for selecting appropriate forecasting models across heterogeneous series and supports adaptive deployment by using the trained forecasters to predict future values for new series.

Empirical results on the PEMS traffic datasets demonstrate consistent gains across multiple horizons, indicating that validation-driven specialization effectively captures heterogeneous predictive structure. Robust Huber-based reassignment further improves stability, yielding larger error reductions and more reliable cluster utilization than non-robust alternatives.

Several limitations merit attention. First, performance depends on the representativeness of the validation split, since reassignment, model selection, and fallback decisions are driven by validation risk. Second, the current framework relies on hard assignments, which may be restrictive when predictive dynamics overlap. Third, results may depend on the choice of candidate cluster sets and the stability of reassignment, particularly under weak heterogeneity or limited sample size. Finally, the approach assumes that one-step training objectives remain informative for multi-horizon validation, which may not always hold.

The framework is model-agnostic and accommodates both point and probabilistic forecasting. Promising directions include extensions to soft or fuzzy specialization \citep{ma2025fcpca}, in which each series is associated with multiple clusters through membership weights, allowing forecasts to be formed as weighted combinations of cluster-specific predictors. These approaches could be further combined with robust assignment schemes \citep{d2021trimmed,ma2025robust}. Another direction is integration with distribution-free uncertainty quantification methods such as conformal prediction \citep{stocker2025gentle}, enabling calibrated predictive intervals under heterogeneous dynamics. In addition, the validation-based assignment rule suggests a natural pathway toward adaptive deployment in evolving environments, where newly observed series can be routed to suitable models and incorporated into cluster-specific updates over time.

Overall, the proposed framework provides a principled and robust approach to forecasting heterogeneous high-dimensional MTS, offering a practical strategy for adaptive model selection while maintaining stable performance across diverse settings.

\section*{Acknowledgments}
This research was supported by King Abdullah University of Science and Technology (KAUST).

\section*{Disclosure statement}\label{disclosure-statement}
The authors declare that they have no known competing financial
interests or personal relationships that could have appeared to
influence the work reported in this paper.

 \section*{Data Availability Statement}\label{data-availability-statement}
 Data sharing is not applicable to this article, as the datasets used in our paper are already publicly available.

  \section*{Code availability}
  All results are coded in Python and are freely available on the following GitHub repository: 
  \url{https://github.com/arbitraryma/Val_driven_forecasting}. For any questions or inquiries, please contact the corresponding author.


\clearpage
\newpage

\addcontentsline{toc}{chapter}{Bibliography}

\bibliographystyle{biom}
\bibliography{paper}

\newpage
\section*{Supplementary Material} 
This section provides additional details  for the two real-data point forecasting experiments.

\subsection*{Leakage-free learning and decision protocol}
This section describes the leakage-free learning and decision protocol underlying the proposed framework. The procedure separates estimation on TRAIN from decision-making on VAL, and integrates clustering, model selection, and fallback decisions within a unified validation-driven pipeline. The complete procedure is summarized in Algorithm~\ref{alg:forecast_clustering}.

\begin{algorithm}[htbp]
\caption{Validation-driven clustering with leakage-free selection and fallback safeguard}
\label{alg:forecast_clustering}
\setstretch{1.0}
\small
\begin{algorithmic}[1]
\Require MTS \(\{\mathbf{X}_i\}_{i=1}^N\); candidate cluster set \(\mathcal{K}\) (or fixed \(K\)); window length \(w\); assignment horizon set \(\mathcal{H}_a\); loss \(\ell\) (Huber \(\ell_\delta\) for point forecasting; pinball \(\rho_q\) for quantile forecasting); warm-start weight \(\eta\); \(K\)-penalty weight \(\gamma\); random seeds \(\mathcal{S}\); maximum outer iterations \(L\).
\Ensure Final assignments \(\{c_i\}\); refit GLOBAL model \(\Theta_g^\star\); refit specialized prototypes \(\{\Theta_k^\star\}_{k=1}^{K^\star}\); TEST scores.

\Statex \textbf{Step 0: Split and preprocess (TRAIN-only statistics)}
\State Split each \(\mathbf{X}_i\) into \((\mathbf{X}_i^{\mathrm{tr}},\mathbf{X}_i^{\mathrm{va}},\mathbf{X}_i^{\mathrm{te}})\).
\State Compute preprocessing statistics on TRAIN only and apply them to TRAIN, VAL, and TEST.

\Statex \textbf{Step 1: Fit GLOBAL on TRAIN}
\State Fit the pooled GLOBAL forecaster: \(\Theta_g \leftarrow \arg\min_{\Theta}\sum_{i=1}^N \mathcal{L}_i^{\mathrm{tr}}(1;\Theta)\).

\Statex \textbf{Step 2: (Optional) Select \(K\) using VAL only}
\If{a candidate set \(\mathcal{K}\) is provided}
    \ForAll{\(K\in\mathcal{K}\)}
        \ForAll{\(s\in\mathcal{S}\)}
            \State Initialize \(\{c_i\}\) into \(K\) clusters using seed \(s\).
            \State Run Steps 3--5; compute \(\mathrm{SelPen}_{\mathrm{VAL}}(K,s)=\mathrm{Sel}_{\mathrm{VAL}}(K,s)+\gamma K/N\).
        \EndFor
        \State Select best seed: \(s_K^\star \leftarrow \arg\min_{s\in\mathcal{S}} \mathrm{SelPen}_{\mathrm{VAL}}(K,s)\).
    \EndFor
    \State Choose \(K^\star \leftarrow \arg\min_{K\in\mathcal{K}} \mathrm{SelPen}_{\mathrm{VAL}}(K,s_K^\star)\).
\Else
    \State Set \(K^\star \leftarrow K\).
\EndIf

\Statex \textbf{Step 3: Initialize assignments}
\State Initialize \(\{c_i\}_{i=1}^N\) into \(K^\star\) clusters.

\Statex \textbf{Step 4: Outer loop (TRAIN fitting + VAL reassignment)}
\For{\(m=1,\ldots,L\)}
    \For{\(k=1,\ldots,K^\star\)}
        \State \(I_k \leftarrow \{i:c_i=k\}\).
        \State Fit prototype \(k\): \(\Theta_k \leftarrow \arg\min_{\Theta}\big[\sum_{i\in I_k}\mathcal{L}_i^{\mathrm{tr}}(1;\Theta)+\eta \|\Theta-\Theta_g\|_2^2\big]\).
    \EndFor
    \State Store previous assignments \(\{c_i^{\mathrm{old}}\}\leftarrow\{c_i\}\).
    \For{\(i=1,\ldots,N\)}
        \For{\(k=1,\ldots,K^\star\)}
            \State \(C_{ik}\leftarrow \frac{1}{|\mathcal{H}_a|}\sum_{h\in\mathcal{H}_a}\mathcal{L}_i^{\mathrm{va}}(h;\Theta_k)\).
        \EndFor
        \State \(c_i \leftarrow \arg\min_k C_{ik}\).
    \EndFor
    \If{\(\{c_i\}=\{c_i^{\mathrm{old}}\}\)}
        \State \textbf{break}
    \EndIf
\EndFor

\Statex \textbf{Step 5: Fallback safeguard (VAL-based, frozen before TEST)}
\For{\(k=1,\ldots,K^\star\)}
    \State \(I_k \leftarrow \{i:c_i=k\}\).
    \State \(L_k^{\mathrm{clus}}\leftarrow \frac{1}{|I_k|}\sum_{i\in I_k}\mathcal{L}_i^{\mathrm{va}}(1;\Theta_k)\),
    \quad \(L_k^{\mathrm{glob}}\leftarrow \frac{1}{|I_k|}\sum_{i\in I_k}\mathcal{L}_i^{\mathrm{va}}(1;\Theta_g)\).
    \State Mark cluster \(k\) as non-specializable if \(L_k^{\mathrm{clus}} > L_k^{\mathrm{glob}}\).
\EndFor
\State Define \(\mathrm{Sel}_{\mathrm{VAL}}(K^\star,s)\) from routed validation loss.

\Statex \textbf{Step 6: Refit on TRAIN+VAL}
\State Refit \(\Theta_g^\star\) on TRAIN+VAL.
\State Refit \(\Theta_k^\star\) on TRAIN+VAL using the GLOBAL model as reference.

\Statex \textbf{Step 7: TEST evaluation (used once)}
\State Evaluate on TEST, using GLOBAL for non-specializable clusters and prototypes otherwise.
\end{algorithmic}
\end{algorithm}

\clearpage
\newpage
\subsection*{Supplementary results for PEMS-BAY} \label{Sup_bay_data}

Tables~\ref{tab:pemsbay_val_select_feat}, \ref{tab:pemsbay_val_select_rand} and \ref{tab:pemsbay_val_select_ours} summarize the validation-based model selection statistics across different numbers of clusters \(K\) for the three clustering strategies.

The results indicate that the optimal number of clusters varies across methods. For the \textsc{Feat-Kmeans} baseline, the best penalized score is obtained at \(K=9\), while for \textsc{Random-Balanced} the optimal choice is \(K=5\). In contrast, our method selects \(K=7\), which is used in the subsequent TEST evaluation. All reported values are scaled by a factor of \(100\).

\begin{table}[htbp]
\centering
 
\caption{Validation selection statistics across random seeds for the feature-based K-means baseline \textsc{Feat-Kmeans} on PEMS-BAY.}
\label{tab:pemsbay_val_select_feat}
\begin{tabular}{c c c c}
\toprule
\(K\) & SelAbs\(_{\mathrm{VAL}}\) (mean\(\pm\)sd) & Best SelAbs\(_{\mathrm{VAL}}\) & Best + Pen \\
\midrule
2 & 12.81 \(\pm\) 0.07 & 12.76 & 12.81 \\
3 & 12.75 \(\pm\) 0.04 & 12.71 & 12.79 \\
4 & 12.62 \(\pm\) 0.04 & 12.56 & 12.67 \\
5 & 12.37 \(\pm\) 0.30 & 12.01 & 12.15 \\
6 & 12.09 \(\pm\) 0.07 & 11.98 & 12.14 \\
7 & 11.99 \(\pm\) 0.05 & 11.95 & 12.14 \\
8 & 11.99 \(\pm\) 0.05 & 11.91 & 12.13 \\
\textbf{9} & \textbf{11.94 \(\pm\) 0.07} & \textbf{11.85} & \textbf{12.10\(^{\star}\)} \\
\bottomrule
\end{tabular}
\end{table}

\begin{table}[htbp]
\centering
 
\caption{Validation selection statistics across random seeds for the Random-Balanced baseline \textsc{Random-Balanced} on PEMS-BAY.}
\label{tab:pemsbay_val_select_rand}
\begin{tabular}{c c c c}
\toprule
\(K\) & SelAbs\(_{\mathrm{VAL}}\) (mean\(\pm\)sd) & Best SelAbs\(_{\mathrm{VAL}}\) & Best + Pen \\
\midrule
2 & 12.65 \(\pm\) 0.09 & 12.53 & 12.59 \\
3 & 12.48 \(\pm\) 0.23 & 12.22 & 12.30 \\
4 & 12.47 \(\pm\) 0.19 & 12.14 & 12.25 \\
\textbf{5} & \textbf{12.41 \(\pm\) 0.19} & \textbf{12.06} & \textbf{12.20\(^{\star}\)} \\
6 & 12.32 \(\pm\) 0.11 & 12.14 & 12.30 \\
7 & 12.22 \(\pm\) 0.16 & 12.03 & 12.22 \\
8 & 12.20 \(\pm\) 0.14 & 12.02 & 12.25 \\
9 & 12.29 \(\pm\) 0.12 & 12.12 & 12.37 \\
\bottomrule
\end{tabular}
\end{table}

\begin{table}[htbp]
\centering
 
\caption{Validation selection statistics across random seeds for the proposed VAL-driven method on PEMS-BAY. Selection uses routed VAL Huber loss with penalty.}
\label{tab:pemsbay_val_select_ours}
\begin{tabular}{c c c c}
\toprule
\(K\) & SelAbs\(_{\mathrm{VAL}}\) (mean\(\pm\)sd) & Best SelAbs\(_{\mathrm{VAL}}\) & Best + Pen \\
\midrule
2 & 4.32 \(\pm\) 0.10 & 4.23 & 4.28 \\
3 & 4.16 \(\pm\) 0.13 & 4.01 & 4.09 \\
4 & 4.18 \(\pm\) 0.06 & 4.11 & 4.22 \\
5 & 4.10 \(\pm\) 0.05 & 4.02 & 4.16 \\
6 & 4.08 \(\pm\) 0.09 & 3.99 & 4.15 \\
\textbf{7} & \textbf{4.02 \(\pm\) 0.08} & \textbf{3.88} & \textbf{4.08\(^{\star}\)} \\
8 & 4.07 \(\pm\) 0.06 & 3.98 & 4.20 \\
9 & 4.11 \(\pm\) 0.10 & 3.99 & 4.24 \\
\bottomrule
\end{tabular}
\end{table}

\clearpage
\newpage
\subsection*{Supplementary results for PEMS-SF}

Tables~\ref{tab:val_select_featkmeans}, \ref{tab:val_select_randombalanced}, and \ref{tab:val_select_ours} summarize the validation-based model selection results across different numbers of clusters \(K\). For the \textsc{Feat-Kmeans} baseline, the best penalized score is obtained at \(K=9\), while for \textsc{Random-Balanced} the optimal choice is \(K=5\), both based on validation MSE. In contrast, our method selects \(K=7\) using validation Huber loss, which is then used in the subsequent TEST evaluation. All reported values are scaled by a factor of \(100\).

\begin{table}[htbp]
\centering
 
\caption{Validation selection statistics across random seeds for the feature-based K-means baseline \textsc{Feat-Kmeans} on PEMS-SF.}
\label{tab:val_select_featkmeans}

\setlength{\tabcolsep}{6pt}
\renewcommand{\arraystretch}{1.15}

\begin{tabular}{c c c c}
\toprule
\(K\) & \(\mathrm{SelAbs}_{\mathrm{VAL}}\) (mean \(\pm\) sd) & Best \(\mathrm{SelAbs}_{\mathrm{VAL}}\) & Best \(+\) Pen \\
\midrule
2 & 85.83 \(\pm\) 0.15 & 85.72 & 85.75 \\
3 & 86.44 \(\pm\) 0.15 & 86.24 & 86.28 \\
4 & 85.20 \(\pm\) 0.46 & 84.59 & 84.59 \\
5 & 84.22 \(\pm\) 0.26 & 84.05 & 84.05 \\
\textbf{6} & \textbf{84.12 \(\pm\) 0.37} & \textbf{83.72} & \textbf{83.72}\(^{\star}\) \\
7 & 84.18 \(\pm\) 0.21 & 83.94 & 83.94 \\
8 & 84.10 \(\pm\) 0.19 & 83.92 & 83.92 \\
9 & 84.06 \(\pm\) 0.21 & 83.96 & 83.96 \\
\bottomrule
\end{tabular}
\end{table}

\begin{table}[htbp]
\centering
\caption{Validation selection statistics across random seeds for the random balanced baseline \textsc{Random-Balanced} on PEMS-SF.}
\label{tab:val_select_randombalanced}
 
\setlength{\tabcolsep}{6pt}
\renewcommand{\arraystretch}{1.15}

\begin{tabular}{c c c c}
\toprule
\(K\) & \(\mathrm{SelAbs}_{\mathrm{VAL}}\) (mean \(\pm\) sd) & Best \(\mathrm{SelAbs}_{\mathrm{VAL}}\) & Best \(+\) Pen \\
\midrule
2 & 86.96 \(\pm\) 0.04 & 86.92 & 86.92 \\
3 & 86.93 \(\pm\) 0.04 & 86.93 & 86.93 \\
4 & 86.90 \(\pm\) 0.06 & 86.86 & 86.86 \\
5 & 86.88 \(\pm\) 0.06 & 86.84 & 86.84 \\
6 & 87.16 \(\pm\) 0.05 & 87.19 & 87.19 \\
7 & 86.88 \(\pm\) 0.06 & 86.86 & 86.86 \\
8 & 86.80 \(\pm\) 0.06 & 86.80 & 86.80 \\
\textbf{9} & \textbf{86.76 \(\pm\) 0.06} & \textbf{86.79} & \textbf{86.79}\(^{\star}\) \\
\bottomrule
\end{tabular}
\end{table}

\begin{table}[H]
\centering
\caption{Validation selection statistics across random seeds for the VAL-driven clustering method (\textsc{Ours}) on PEMS-SF.}
\label{tab:val_select_ours}
 
\setlength{\tabcolsep}{6pt}
\renewcommand{\arraystretch}{1.15}

\begin{tabular}{c c c c}
\toprule
\(K\) & \(\mathrm{SelAbs}_{\mathrm{VAL}}\) (mean \(\pm\) sd) & Best \(\mathrm{SelAbs}_{\mathrm{VAL}}\) & Best \(+\) Pen \\
\midrule
2 & 21.12 \(\pm\) 0.14 & 21.02 & 21.08 \\
3 & 20.51 \(\pm\) 0.13 & 20.35 & 20.38 \\
4 & 20.45 \(\pm\) 0.06 & 20.36 & 20.41 \\
5 & 20.41 \(\pm\) 0.09 & 20.36 & 20.42 \\
6 & 20.39 \(\pm\) 0.09 & 20.30 & 20.36 \\
\textbf{7} & \textbf{20.35 \(\pm\) 0.11} & \textbf{20.19} & \textbf{20.27}\(^{\star}\) \\
8 & 20.36 \(\pm\) 0.09 & 20.26 & 20.35 \\
9 & 20.40 \(\pm\) 0.10 & 20.33 & 20.38 \\
\bottomrule
\end{tabular}
\end{table}

\begin{figure}[t]
    \centering
    \includegraphics[width=0.95\linewidth]{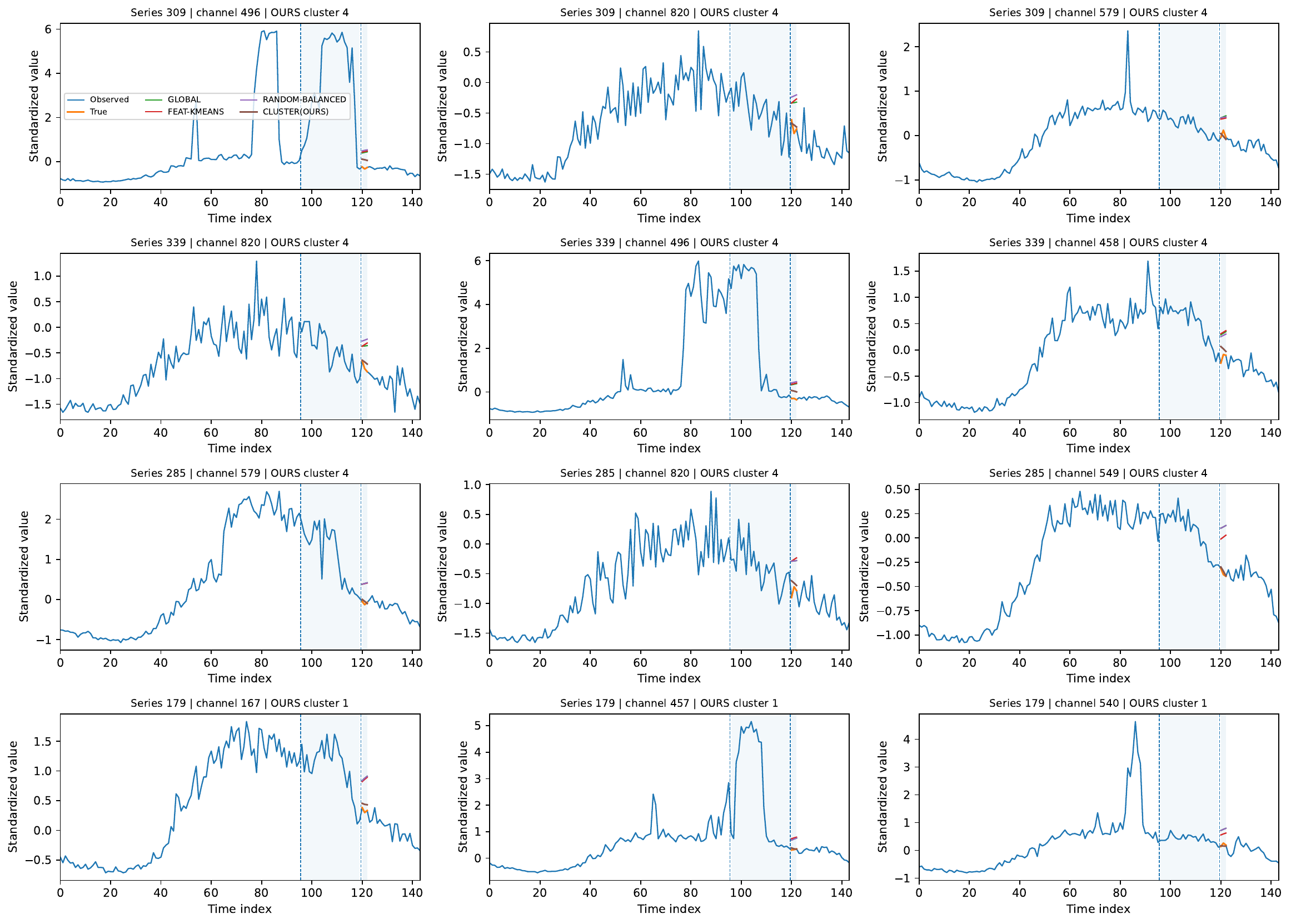}
    \caption{
    {Representative trajectory-level comparisons at horizon \(h=3\).}
    Each subplot compares GLOBAL, \textsc{Feat-Kmeans}, \textsc{Random-Balanced}, and \textsc{Cluster(Ours)} against the true future trajectory for selected series--channel pairs.
    \textsc{Cluster(Ours)} generally remains closer to the held-out future trajectory, illustrating consistent trajectory-level improvements.
    }
    \label{fig:rep_panels_h3}
\end{figure}

\begin{figure}[t]
    \centering
    \includegraphics[width=0.95\linewidth]{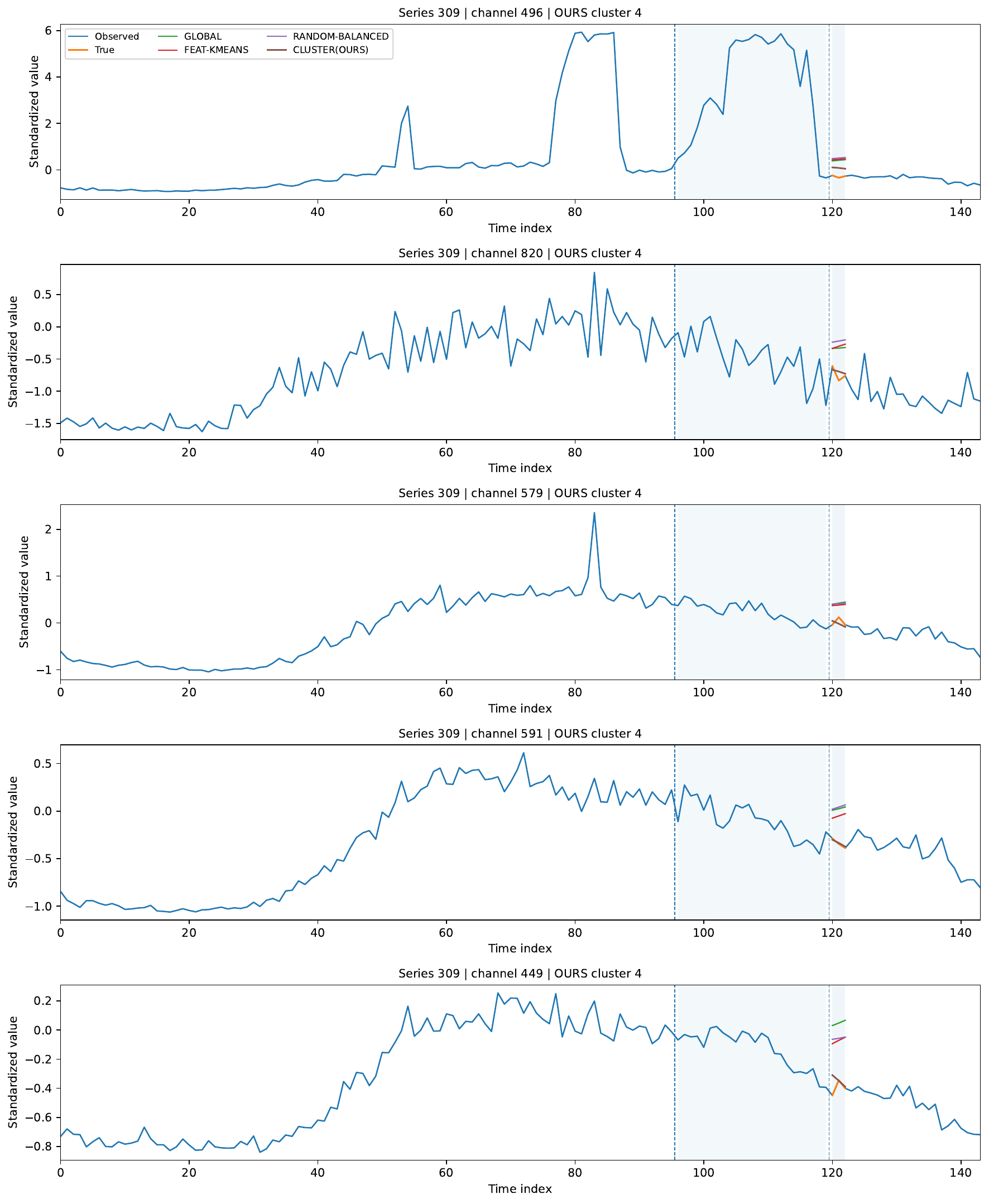}
    \caption{
    {Single-series full-day prediction comparison at horizon \(h=3\).}
    The dashed vertical lines indicate the validation and test split points, and shaded regions correspond to the validation and test segments.
    The forecasts from \textsc{Cluster(Ours)} more closely follow the held-out future trajectory across the selected channels.
    }
    \label{fig:single_series_h3}
\end{figure}

Figure~\ref{fig:single_series_h6} provides trajectory-level evidence at horizon \(h=6\). Figures~\ref{fig:rep_panels_h3} and \ref{fig:single_series_h3} provide additional trajectory-level evidence at horizon \(h=3\). Overall, \textsc{Cluster(Ours)} remains closer to the held-out future trajectory than GLOBAL, \textsc{Feat-Kmeans}, and \textsc{Random-Balanced}. These results are consistent with the aggregate performance comparisons and illustrate that the proposed method maintains stable improvements even in shorter-horizon settings.

\end{document}